\documentclass[fleqn,usenatbib]{mnras}

\usepackage{amsmath}
\usepackage[T1]{fontenc}

\DeclareRobustCommand{\VAN}[3]{#2}
\let\VANthebibliography\thebibliography
\def\thebibliography{\DeclareRobustCommand{\VAN}[3]{##3}\VANthebibliography}

\usepackage{graphicx}	% Including figure files
\usepackage{amsmath}	% Advanced maths commands
\usepackage{amssymb}	% Extra maths symbols
\usepackage{newtxtext,newtxmath}

\newcommand{\h}{^{\textrm h}}
\newcommand{\m}{^{\textrm m}}
\newcommand{\s}{^{\textrm s}}

\title[Discovery of 13 new pulsars in $\omega$-Centauri]{MeerKAT discovery of 13 new pulsars in Omega Centauri}

\author[W. Chen et al.]{
W.~Chen,$^{1}$\thanks{E-mail: wchen@mpifr-bonn.mpg.de}
P.~C.~C.~Freire,$^{1}$
A.~Ridolfi,$^{3,1}$
E.~D.~Barr,$^{1}$
B.~Stappers,$^{2}$
M.~Kramer,$^{1,2}$
A.~Possenti,$^{3}$
\and
S.~M.~Ransom,$^{4}$
L.~Levin,$^{2}$
R.~P.~Breton,$^{2}$
M.~Burgay,$^{3}$
F.~Camilo,$^{5}$
S.~Buchner,$^{5}$
D.~J.~Champion,$^{1}$
\and
F.~Abbate,$^{1}$
V.~Venkatraman~Krishnan,$^{1}$
P.~V.~Padmanabh,$^{1,8,9}$
T.~Gautam,$^{1}$
L.~Vleeschower,$^{2}$
\and
M.~Geyer,$^{5}$
J-M.~Grie{\ss}meier,$^{6,7}$
Y.~P.~Men,$^{1}$
V.~Balakrishnan,$^{1}$
M.~C.~Bezuidenhout$^{2}$
\\
$^{1}$Max-Planck-Institut f\"{u}r Radioastronomie, Auf dem H\"{u}gel 69, D-53121 Bonn, Germany\\
$^{2}$Jodrell Bank Centre for Astrophysics, Department of Physics and Astronomy, The University of Manchester, Manchester M13 9PL, UK\\
$^{3}$INAF -- Osservatorio Astronomico di Cagliari, Via della Scienza 5, I-09047 Selargius (CA), Italy\\
$^{4}$National Radio Astronomy Observatory (NRAO), 520 Edgemont Rd., Charlottesville, VA 22903 USA\\
$^{5}$South African Radio Astronomy Observatory (SARAO), 2 Fir Street, Black River Park, Observatory, Cape Town 7925, South Africa\\
$^{6}$LPC2E - Universit\'{e} d'Orl\'{e}ans / CNRS, 45071 Orl\'{e}ans cedex 2, France\\
$^{7}$Observatoire Radioastronomique de Nançay (ORN), Observatoire de Paris, Universit\'{e} PSL, Univ Orl\'{e}ans, CNRS, 18330 Nan{c}ay, France\\
$^{8}$ Max Planck Institute for Gravitational Physics (Albert Einstein Institute), D-30167 Hannover, Germany\\
$^{9}$ Leibniz Universit\"{a}t Hannover, D-30167 Hannover, Germany\\
}
\date{Accepted XXX. Received YYY; in original form ZZZ}

\pubyear{2015}

\begin{document}
\label{firstpage}
\pagerange{\pageref{firstpage}--\pageref{lastpage}}
\maketitle

\begin{abstract}
The most massive globular cluster in our Galaxy, Omega Centauri, is an interesting target for pulsar searches, because of its multiple stellar populations and the intriguing possibility that it was once the nucleus of a galaxy that was absorbed into the Milky Way. The recent discoveries of pulsars in this globular cluster and their association with known X-ray sources was a hint that, given the large number of known X-ray sources, there is a much larger undiscovered pulsar population. We used the superior sensitivity of the MeerKAT radio telescope to search for pulsars in Omega Centauri. In this paper, we present some of the first results of this survey, including the discovery of 13 new pulsars; the total number of known pulsars in this cluster currently stands at 18. At least half of them are in binary systems and preliminary orbital constraints suggest that most of the binaries have light companions. We also discuss the ratio between isolated and binaries pulsars and how they were formed in this cluster.
\end{abstract}


%%%%%%%%%%%%%%%%%%%%%%%%%%%%%%%%%%%%%%%%%%%%%%%%%%

%%%%%%%%%%%%%%%%% BODY OF PAPER %%%%%%%%%%%%%%%%%%

\section{Introduction}

% Globular cluster is significant because ...

Pulsar surveys conducted in Globular clusters (GCs) have yielded fruitful rewards in recent decades, with the discovery of a total of 261 pulsars\footnote{\url{https://www3.mpifr-bonn.mpg.de/staff/pfreire/GCpsr.html}}. Per unit of stellar mass, GC have three orders of magnitude more pulsars than the Galactic disk. The reason for this is their large stellar densities: these prompt stellar interactions \citep{Verbunt1987} where old, dead Neutron stars (NSs) acquire new main sequence (MS) companions. These then evolve, forming X-ray binaries (also exceptionally numerous in GCs, \citealt{Clark1975}), where the NS is being spun up by accretion of matter from the MS star. When the accretion stops, these systems become millisecond pulsar (MSP) binaries, which have nearly circular orbits and low-mass companions \citep{Bhattacharya1991}. Indeed, the pulsar population in GCs is dominated by such binaries. 

However, in some cases, additional exchange encounters can originate different end products. If they happen during the X-ray binary phase, they can disrupt the binary and produce many single and/or partially recycled pulsars, which are not only slower, but appear to be much younger than the GC population \citep{Verbunt2014}.
Such encounters can also replace a pulsar's companion by a much more massive degenerate object, resulting in massive, eccentric binary MSPs unlike any seen in the Galactic disk (e.g., \citealt{Freire2004,Lynch2012,DeCesar2015,Ridolfi2021,Ridolfi2022}). If the massive companions happen to be stellar-mass black holes, these systems could be superb test-beds for fundamental physics \citep{Liu2014}.

Such exotic pulsar binaries are generally observed in GCs with very dense cores, especially core-collapsed clusters; these are the types of environments where each particular binary is likely to go through more than one disruptive stellar encounter \citep{Verbunt2014}. Thus, the pulsar population in a GC, and the types of binaries the pulsars find themselves in, reflects not only its current dynamical status, but also the cluster's previous evolution \citep{Benacquista2013}.

\subsection{The Omega Centauri globular cluster}

Omega Centauri ($\omega$-Cen, also known as NGC 5139), the largest GC in our Galaxy, is a natural target to search for pulsars. It is located in the constellation of Centaurus and is 5.2 kpc away from the Sun, with an age of 11.52 Gyr \citep{Forbes2010}. Besides its size and large number of stars, it differs from other GCs because of its intricate composition of different populations of stars \citep{Bedin2004}. This could indicate that $\omega$-Cen is the merger of several clusters, like the Sagittarius dwarf galaxy \citep{Ibata1994}. Moreover, it was found to be rich in calcium and heavy metals \citep{Lee2009}, which is a tracer of supernovae explosions. However these materials ejected by the explosion could not be sustained by the current gravitational potential of the cluster. This along with the multiple stellar populations supports the long established idea that $\omega$-Cen is the relic of a former disrupted dwarf galaxy \citep{Hilker2000, Ibata2019}.

% previously searches
Unassociated high-energy emission in GCs is thought to originate from MSPs \citep{Venter2009} as supported by observations, (e.g. \citealt{Abdo2009}). $\gamma$-ray emission has been detected in several GCs by the Fermi Large Area Telescope \citep{Abdo2010}, including those that, at the time, had no previously known pulsars, such as $\omega$-Cen.

Additionally, $\sim$30 unassociated X-ray sources have been found within the core of $\omega$-Cen to have luminosities similar to the emission of pulsars in other GCs \citep{Henleywillis2018}. However, previous searches for pulsars in this cluster turned out to be unsuccessful \citep{Edwards2001, Possenti2005, Camilo2015}.

Notwithstanding, in 2019, \cite{Dai2020} carried out a search for pulsars using the new Ultra-wide Bandwidth receiver \citep[UWL,][]{Hobbs2020} of the 64-m "Murriyang" radio telescope at Parkes, NSW, Australia. This finally allowed the discovery of the first 5 MSPs in $\omega$-Cen. Among these, PSR J1326$-$4728B is in an eclipsing binary system with a light companion, and it is associated with an X-ray source \citep{Henleywillis2018}. The authors suggested that the non-detection of pulsars from the previous searches on this cluster was caused by the lack of sensitivity of previous surveys. Given the large distance of $\omega$-Cen (and other GCs in general) it is clear that we are only detecting the very brightest pulsars in them. This means that more sensitive telescopes would in principle detect many more pulsars in this and other globular clusters.

\subsection{The MeerKAT survey}

The MeerKAT 64-antenna array, located in the Karoo desert in South Africa, \citep{Jonas2016, Camilo2018} started scientific observations in late 2019, becoming, by far, the most radio sensitive radio telescope in the Southern Hemisphere. Numerous targeted observations and surveys of pulsars have been carried out since and are producing significant results. For example, the TRAnsients and PUlsars with MeerKAT \citep[TRAPUM,][]{StappersKramer2016} Large Survey Project aims to increase the total known population of pulsars, and discover peculiar binary pulsars that might be suitable for studies of fundamental physics and stellar evolution. This project has so far discovered 156 pulsars\footnote{\url{http://www.trapum.org/discoveries/}}. About one-third of these (e.g. \citealt{Ridolfi2021, Douglas2022, Ridolfi2022, Vleeschower2022, Abbate2022}), are found in various GCs, including 47 Tucanae (Chen et al., in prep.), a cluster that has been searched for over 20 years. 

Given the small beams produced by the phased interferometer, pulsars surveys need a beamformer in order to cover as much sky simultaneously as possible. The beam former developed by
\citet{Barr2018} and \citet{Chen2021} can generate up to 1000 coherent beams that cover a significant part of the primary field of view of an individual antenna, with high time and frequency resolution. Such a large number of beams is essential for blind surveys (which cover mostly the Galactic plane), but they are also important for GCs. Indeed, even though GCs have a relatively small angular size on the sky, we still often need hundreds of beams to cover their half-mass radii, which are the regions where most pulsars are likely to be located. This is especially true in the case of $\omega$-Cen, which is the most massive and has the largest projected size among all known GCs.

In this paper, we report the result of the searches for the new pulsars in two observations of $\omega$-Cen made with MeerKAT. Section \ref{sec:observations} describes the observational parameters and analysis of the recorded data. Section \ref{sec:discoveries} presents the new discoveries, while Section \ref{sec:discussion} discusses the uncovered pulsar population and its implications in our understanding of the cluster.

\section{Observations and data analysis}
\label{sec:observations}

\subsection{Observations}
We carried out two multibeam observations of $\omega$-Cen with MeerKAT on 2021 March 21 and 26, as part of the TRAPUM GC survey. The cluster was observed for 4 hours in each observation using the L-band receivers, which cover the 856--1712 MHz frequency range. The coherent beams were synthesised using the Filterbanking BeamFormer User-Supplied Equipment \citep[FBFUSE,][]{Barr2018}. A tiling of 704 coherent beams was generated by \textsc{mosaic}\footnote{\url{https://github.com/wchenastro/Mosaic}} to cover the cluster, as shown in Figure \ref{fig:tiling}. This was centred at the nominal centre of $\omega$-Cen, at equatorial coordinates (J2000) 13:26:47.24, $-$47:28:46.5 and Galactic coordinates 309.10, 14.97 \citep{Harris2010}, covering a circular region of 7.53 arcmin in radius, i.e. roughly twice the size of the half-light radius of the cluster. During both observations, 60 antennas were used. To create a hexagonal tiling, the corresponding synthesised beam shape was approximated by an ellipse whose major and minor axis were 20.46 arcseconds and 9.32 arcseconds at the middle of the observation on March 21. The method and performance of such approximation is discussed in \citet{Chen2021}. The March 26 observation started at almost the same hour angle, so the beam shape is very similar.

For each beam, the observing band was split into 2048 channels and recorded every 153.12 $\mu$s as filterbank search-mode files by Accelerated Pulsar Search User Supplied Equipment (APSUSE) computing cluster. Once the observation was over, the frequency channels were summed in groups of 16 after being incoherently de-dispersed at the dispersion measure (DM) of 97 pc\,cm$^{-3}$, about the average of the previously known 5 pulsars in $\omega$-Cen.
The resulting sub-banded data preserved 128 channels, significantly reducing the total data volume. With this resolution, the intra-channel smearing time is 1.57 ms, at the lowest channel and the upper bound of the search DM.
% 4.15*(115-97)*(0.856**(-2) - (0.856+0.856/128)**(-2))

\subsection{Sensitivity}

We calculate a minimum detectable flux density of S$_{\textrm{min}}$ = 10 $\mu$Jy for our search on a 4-hour observation, following the modified radiometer equation given by \cite{Dewey1985}:

\begin{align}
    S_{\textrm{min}} = \frac{\textrm{S/N}\,\beta\,T_{\textrm{sys}}}{\mathcal {E}_\textrm{FFT}\,G\sqrt{n_{\textrm{pol}}\,B\,\Delta\,t_{\textrm{obs}}}} \sqrt{\frac{\zeta}{1-\zeta}}
\end{align}

where S/N is the minimal signal-to-noise of 10 for a valid detection; $\beta$ is the correction factor of 1.01 to compensate the loss of sensitivity during the digitization process; $T_{\textrm{sys}}$ is the system temperature of about 26 K\footnote{\url{https://skaafrica.atlassian.net/rest/servicedesk/knowledgebase/latest/articles/view/277315585}}, including the contribution from the receiver temperature of 18 K, the sky temperature of about 3.5 K at L-band and the combination of the spillover noise and the atmosphere of about 4.5 K; $\mathcal {E}_\mathrm{FFT}$ is the FFT search efficiency which is 0.7 according to \citet{Morello2020}; G is the combined gain of the array, which is 2.65 K Jy$^{-1}$ when 60 antennas are used; $n_{\textrm{pol}}$ is the number of polarizations, which is 2 while using the total power beamformer in TRAPUM observations; $B$ is the effective receiver bandwidth of about 640 MHz, after removing the channels affected by RFI; $\Delta t_{\textrm{obs}}$ is the integration time which was 4 hours for both observations; $\zeta$ is the pulse's apparent duty cycle, which we consider to be 8\% with the broadening effects from interstellar medium and the instrument. 
% S_min 10 * 1.01 * 26 / ( 0.7 * 2.65 * sqrt(2 * 0.6e9*4*3600)) * sqrt(0.08/(1-0.08))
% T_sky 2.725+1.6*(1.284)**(-2.75)

\subsection{Data reduction and search pipeline}
\label{subsec:data_reduction_and_saerch_pipeline}
The data were transferred to Germany from South Africa in hard drives. Searches for periodic signals were carried out using the Hercules computing cluster\footnote{\url{https://docs.mpcdf.mpg.de/doc/computing/clusters/systems/Radioastronomy.html}}. For this work, we restricted our search to the beams located within the half-light radius of $\omega$-Cen. 
% filtool
The search of an individual beam started with an RFI mitigation procedure using \textsc{filtool}\footnote{\url{https://github.com/ypmen/PulsarX}}. It applies two major operations on the filterbank data according to the statistical analysis. Assuming that the noise follows a Gaussian distribution, it calculates the kurtosis of samples in certain time unit across the full band, to obtain the deviation of each channel then replace the outlying ones with the mean. Similarly, it calculates the skewness of the samples to identify RFIs because they are often non-Gaussian. After this, it normalizes the data across channels so that each channel has an average of 0 and a variance of 1. \par
% PULSAR_MINER
The resulting filterbank data were fed to a pulsar search pipeline supervised by \textsc{pulsar\_miner}\footnote{\url{https://github.com/alex88ridolfi/PULSAR_MINER}}. The pipeline is based on various programs and utilities from \textsc{presto}\footnote{\url{https://github.com/scottransom/presto}} \citep{prestoascl}, a pulsar search and analysis toolkit. It first uses \textsc{rfifind} to examine the data and flag the narrow and wide band interference. The operation outputs a report and a mask for later use. Free electrons in the interstellar medium engaging with passing electromagnetic waves leads to arrival delays across channels, which are quantified as Dispersion Measure (DM). Thus, signals with 0 DM are terrestrial which were identified by \textsc{prepdata} and zapped. After that, the data were split into segments of different lengths, to be sensitive to binary pulsars with different orbital periods \citep{Ransom2003}. The chosen lengths of the segments were 10, 20, 30, 60, 120 minutes, in addition to the full length (4 h) of each observation. Each of these segments were de-dispersed into time series using a number of trial DMs. The step between these DMs were set to 0.1 pc~cm$^{-3}$ which were determined using \texttt{DDplan.py}, a program that generates suitable de-dispersion schemes considering the DM range, central frequency, sampling time, number of channels and other parameters. The final range of DMs searched was 85--115 pc~cm$^{-3}$, which is slightly wider than the range of 90--110 used in \cite{Dai2020}. Each time series was then searched for periodic signals in the Fourier domain using \textsc{accelsearch} \citep{Ransom2002}. In order to account for the drifting of the frequency of a signal due to orbital motion, \textsc{accelsearch} searches for signals that drift linearly over multiple Fourier bins. The maximum number of bins drifted is set by the \texttt{zmax} parameter, which was chosen to be 200 in our search. The candidates from the search went through a sifting process, where only the ones with S/Ns higher than 4 and within the interested DM and period ranges were delivered to the next stage. In the final step, the sifted candidates were folded using \textsc{prepfold} with an extra optimization on the period and DM. The plots of these candidates were inspected by eye.

\section{Discoveries}
\label{sec:discoveries}
\begin{figure*}
    \centering
    \includegraphics[width=0.75\textwidth]{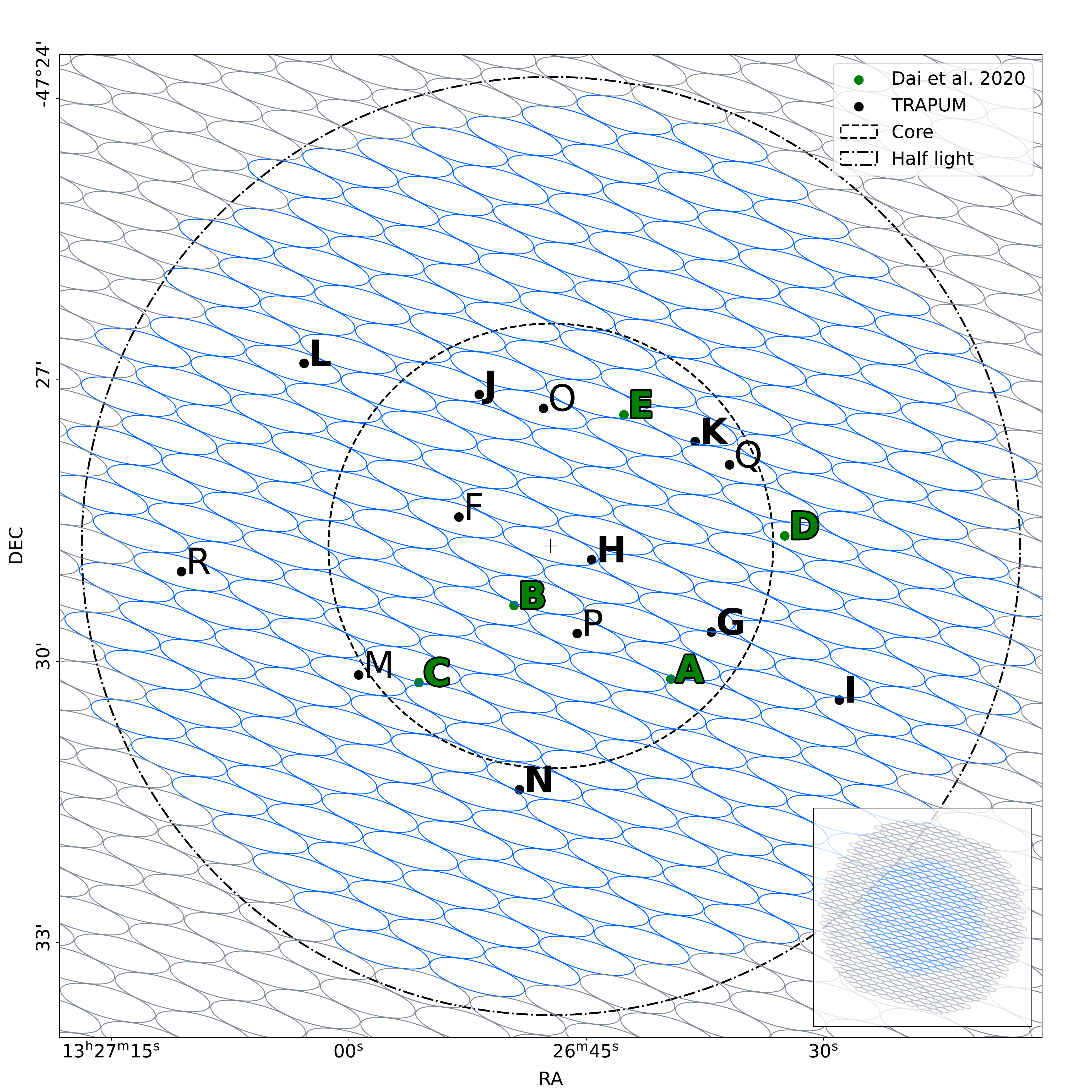}
    \caption{Tiling and detections. Shown is the beam tiling pattern at the start of the observation generated by \textsc{mosaic} on sky, centred at the optical centre (denote with a cross) of $\omega$-Cen. The radii of the core and the half light \citep{Harris2010} are denoted using dashed line and dash-dotted lines. The beams with blue edges were searched for pulsars in this work, roughly cover the region within the half light radius. Lower right is a zoom-out view of the tiling. The position indicated by green with black edges are known sources published in \citep{Dai2020}. The positions indicated by black are the new discoveries in this paper. \textbf{Bold face} indicates that they have timing positions (A and B) or their positions have been constrained using multibeam detections. Others were placed at the centers of the beams where they have the brightest detections.}
    \label{fig:tiling}
\end{figure*}

\begin{figure*}
    \centering
    \setlength\tabcolsep{1.0pt}
    \renewcommand{\arraystretch}{0.1}
    \begin{tabular}{lllllll}
    \includegraphics[width=0.14\textwidth]{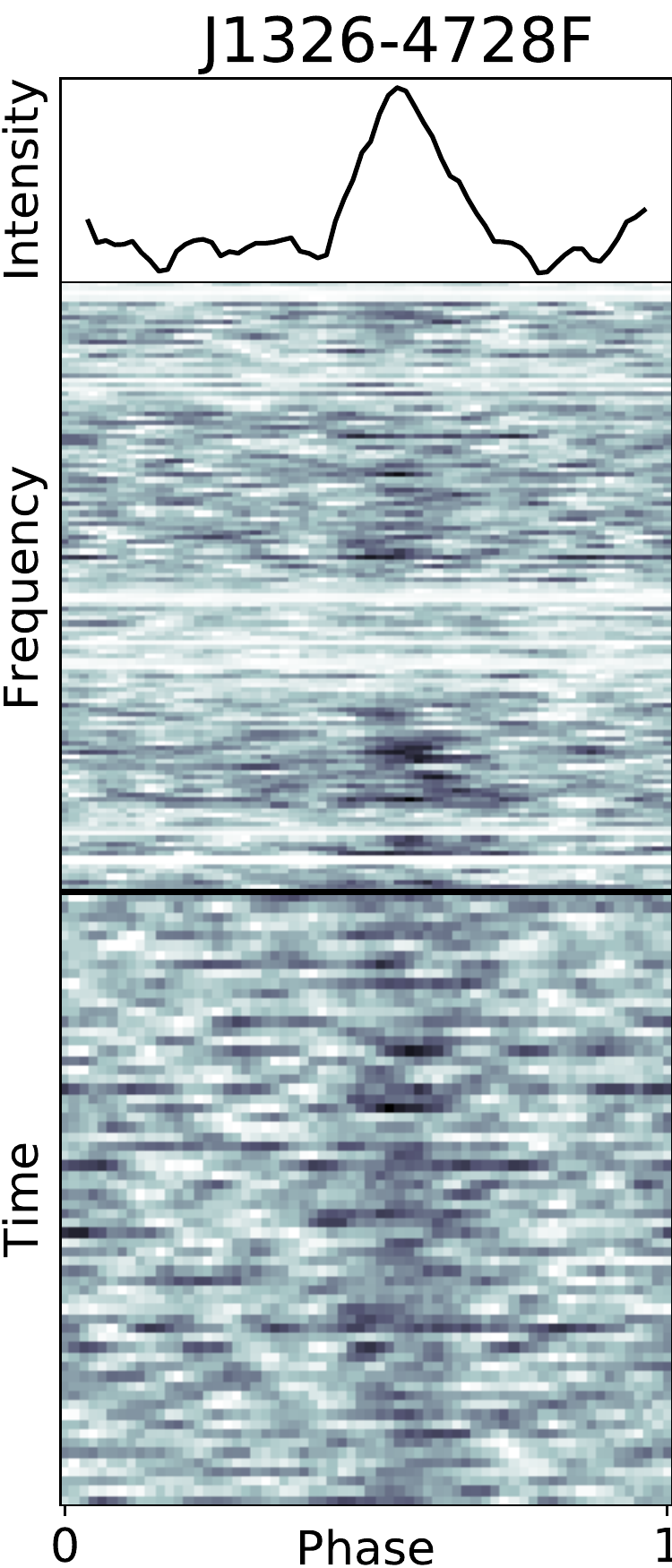} &
    \includegraphics[width=0.14\textwidth]{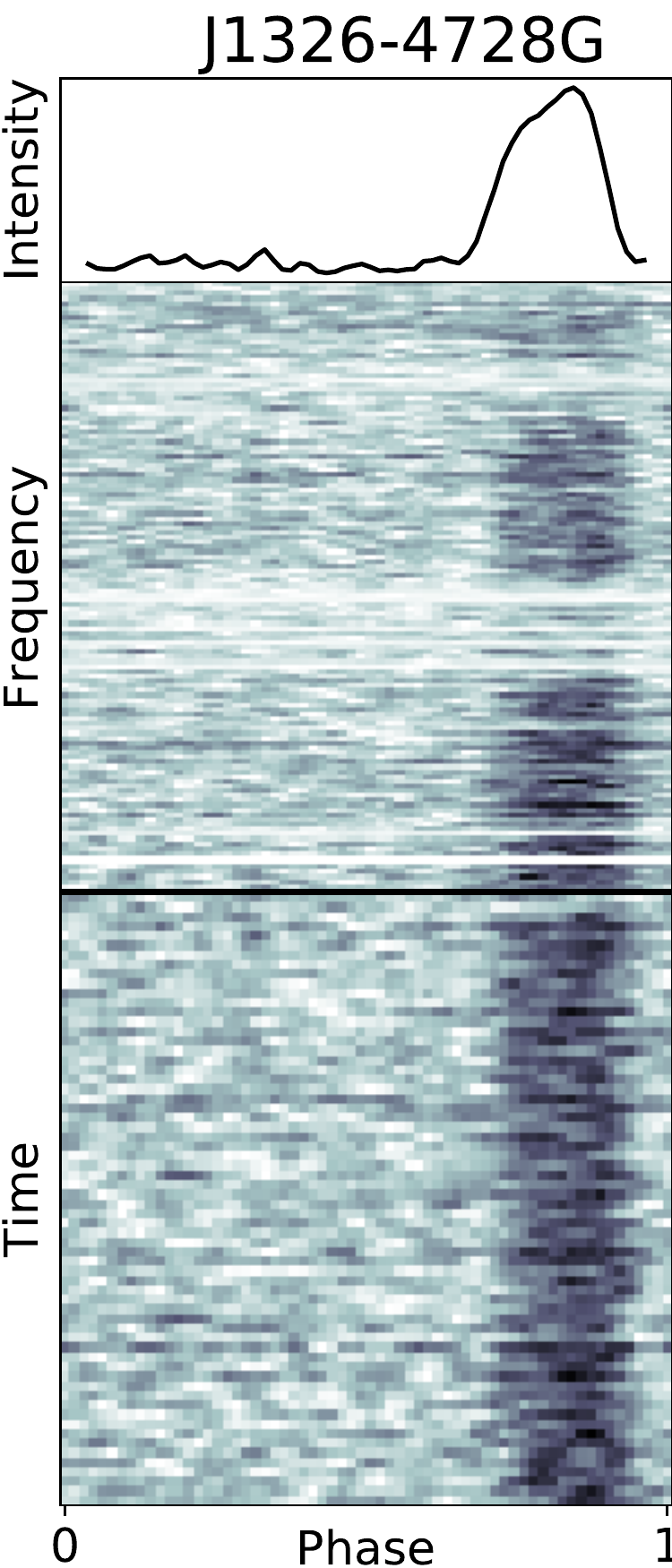} &
    \includegraphics[width=0.14\textwidth]{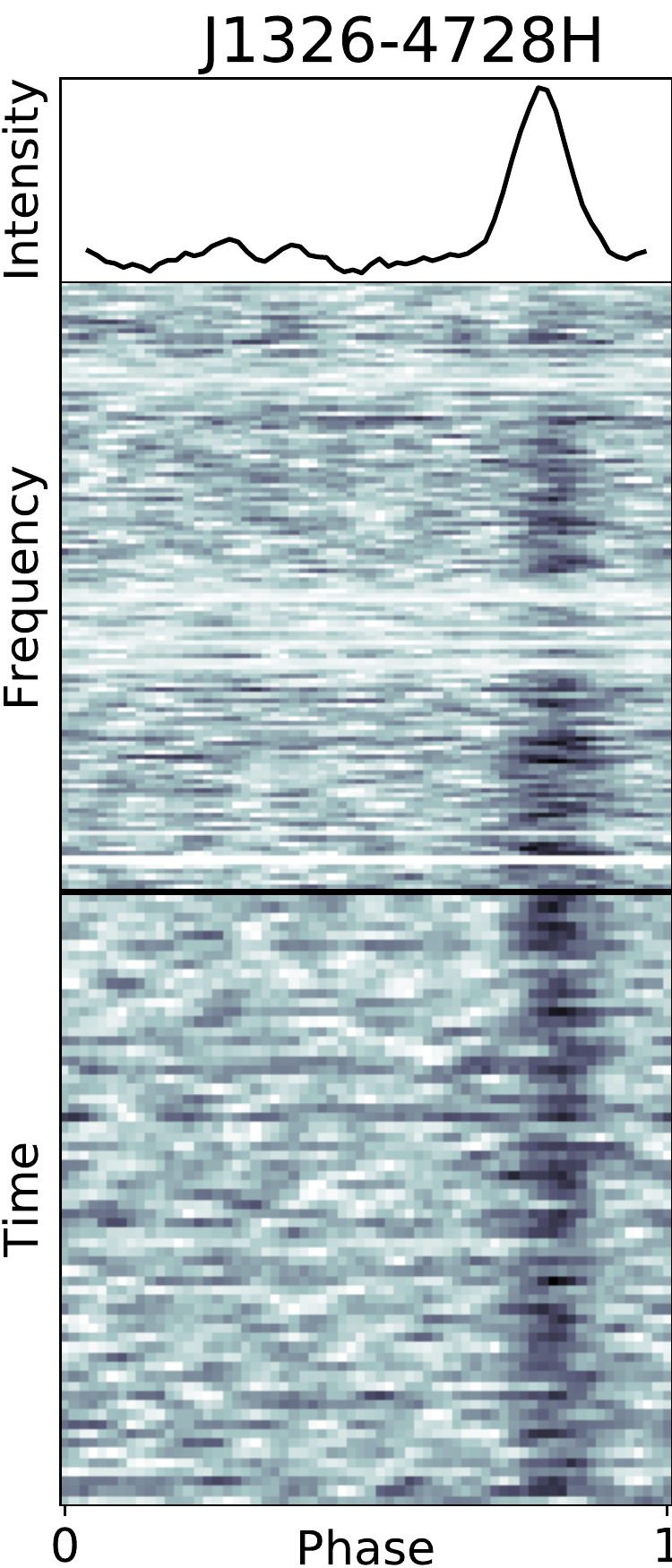} &
    \includegraphics[width=0.14\textwidth]{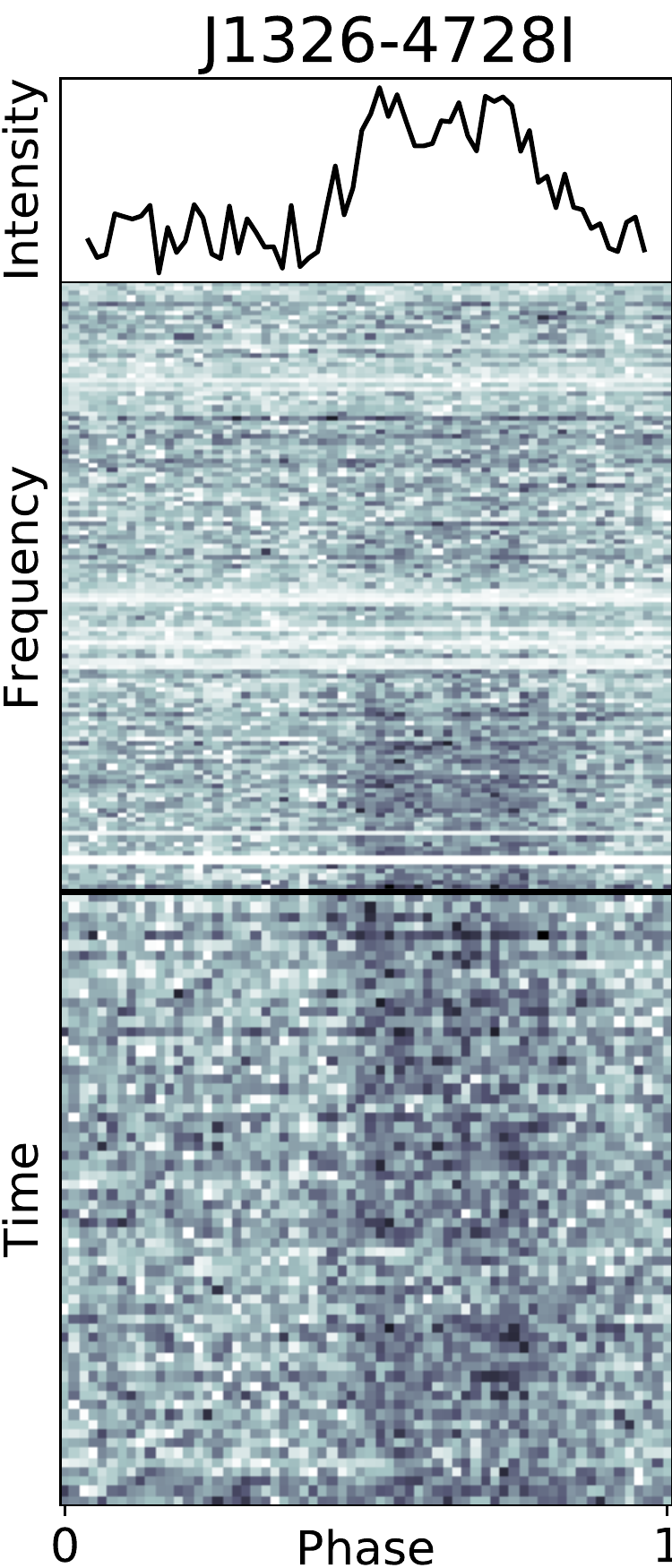} &
    \includegraphics[width=0.14\textwidth]{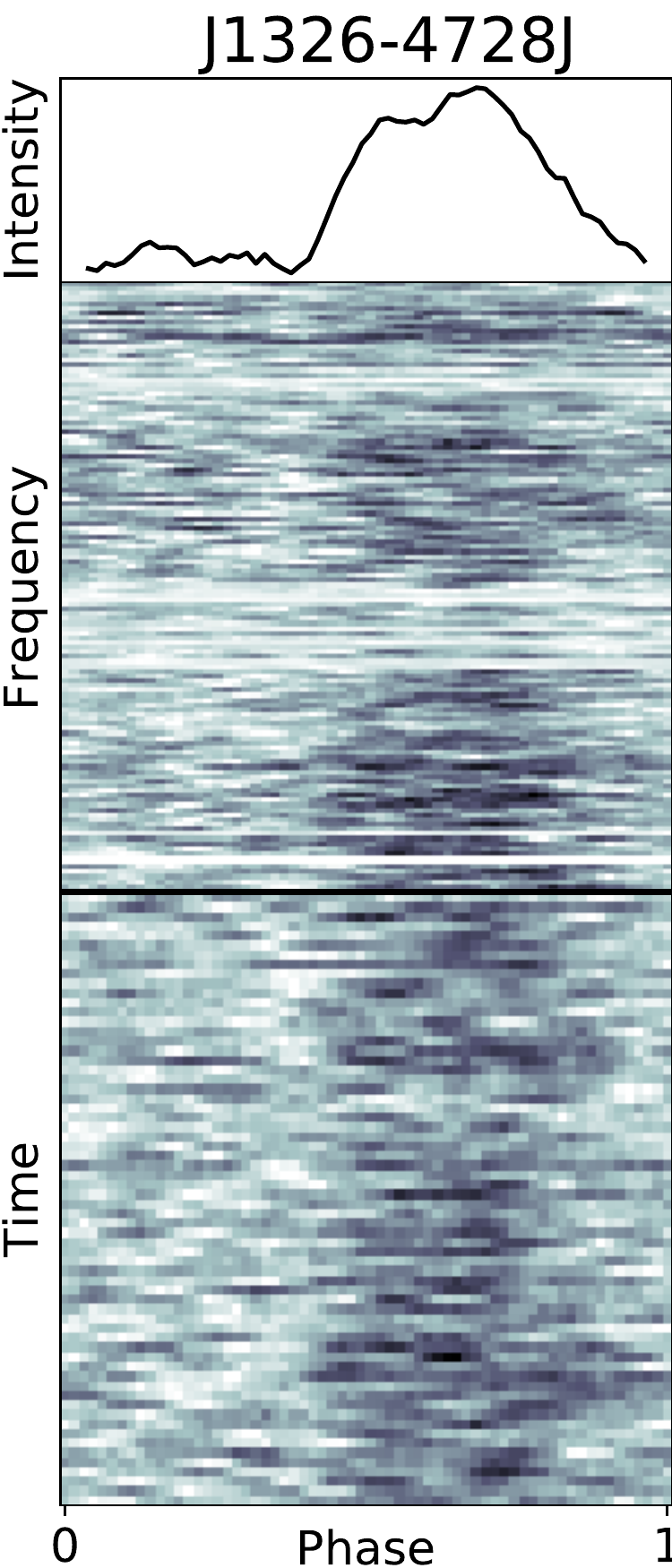} &
    \includegraphics[width=0.14\textwidth]{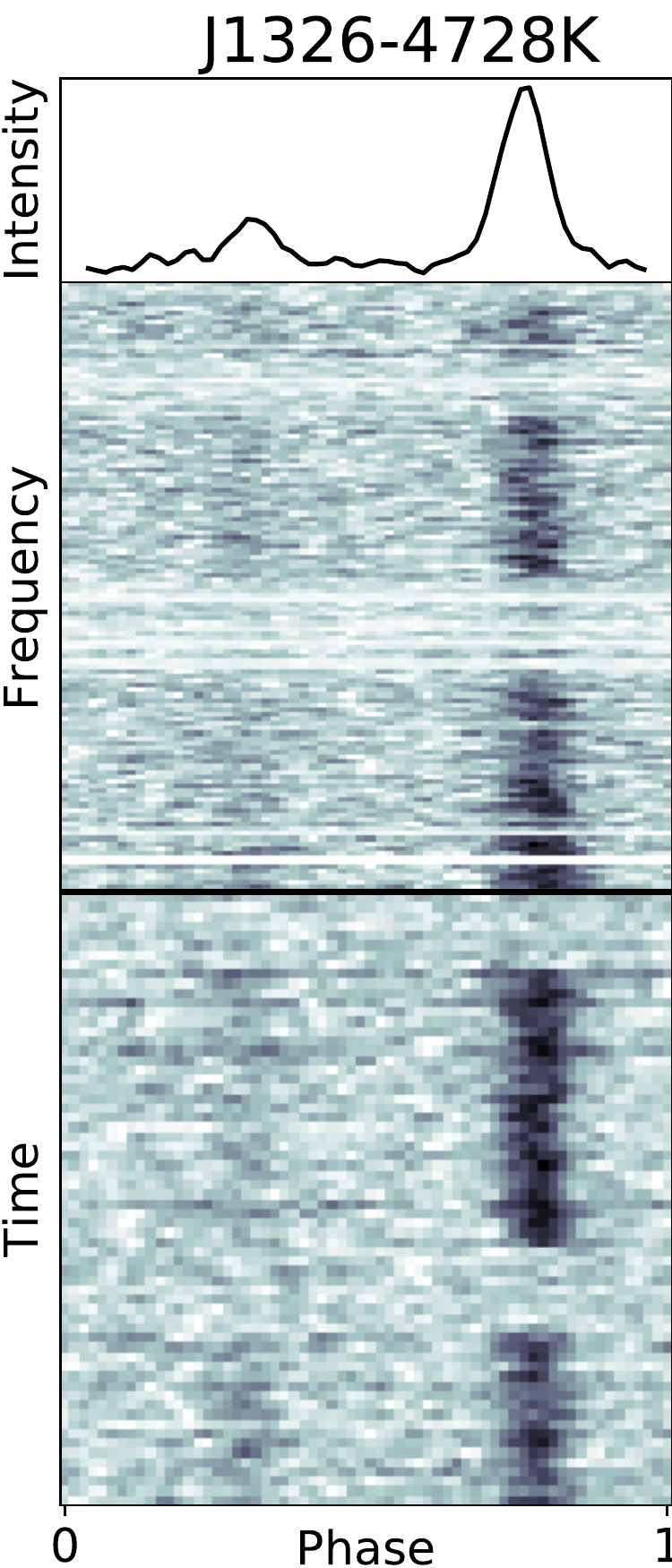} &
    \includegraphics[width=0.14\textwidth]{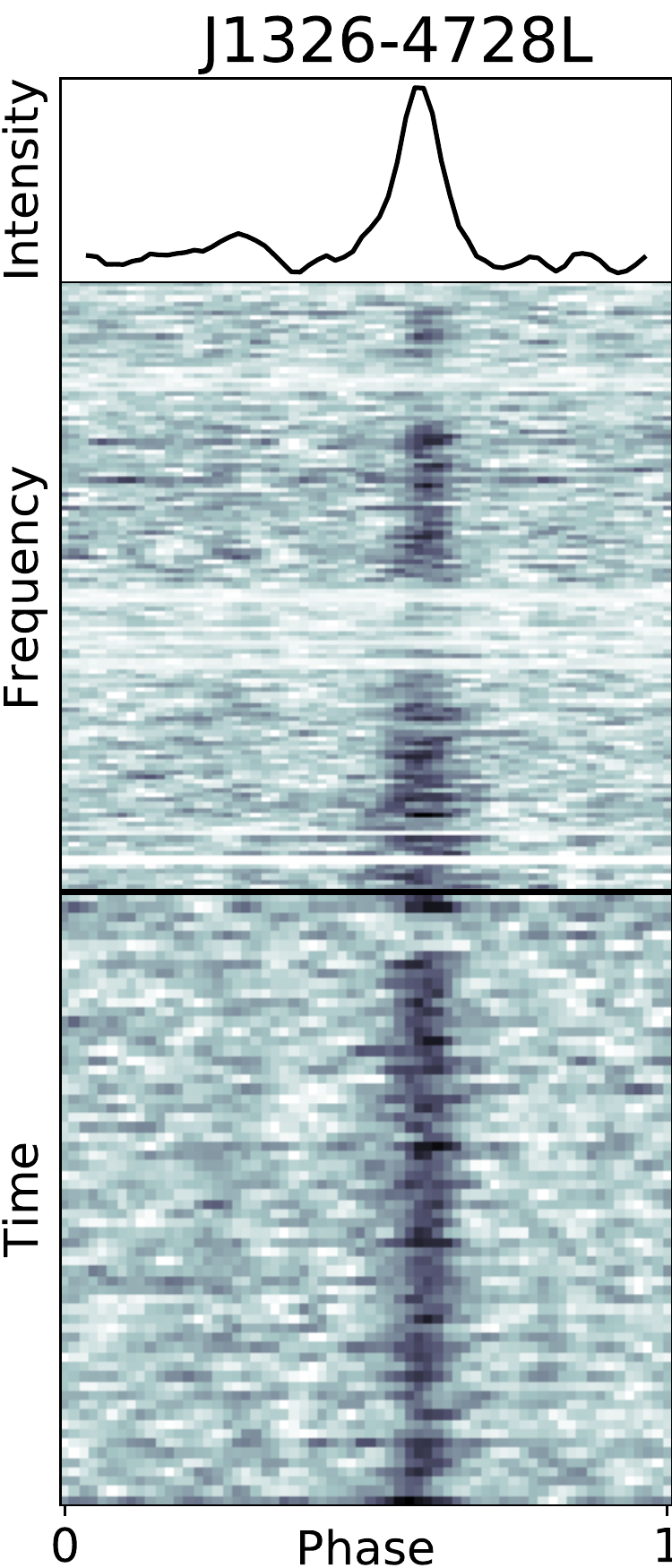} \\
    \includegraphics[width=0.14\textwidth]{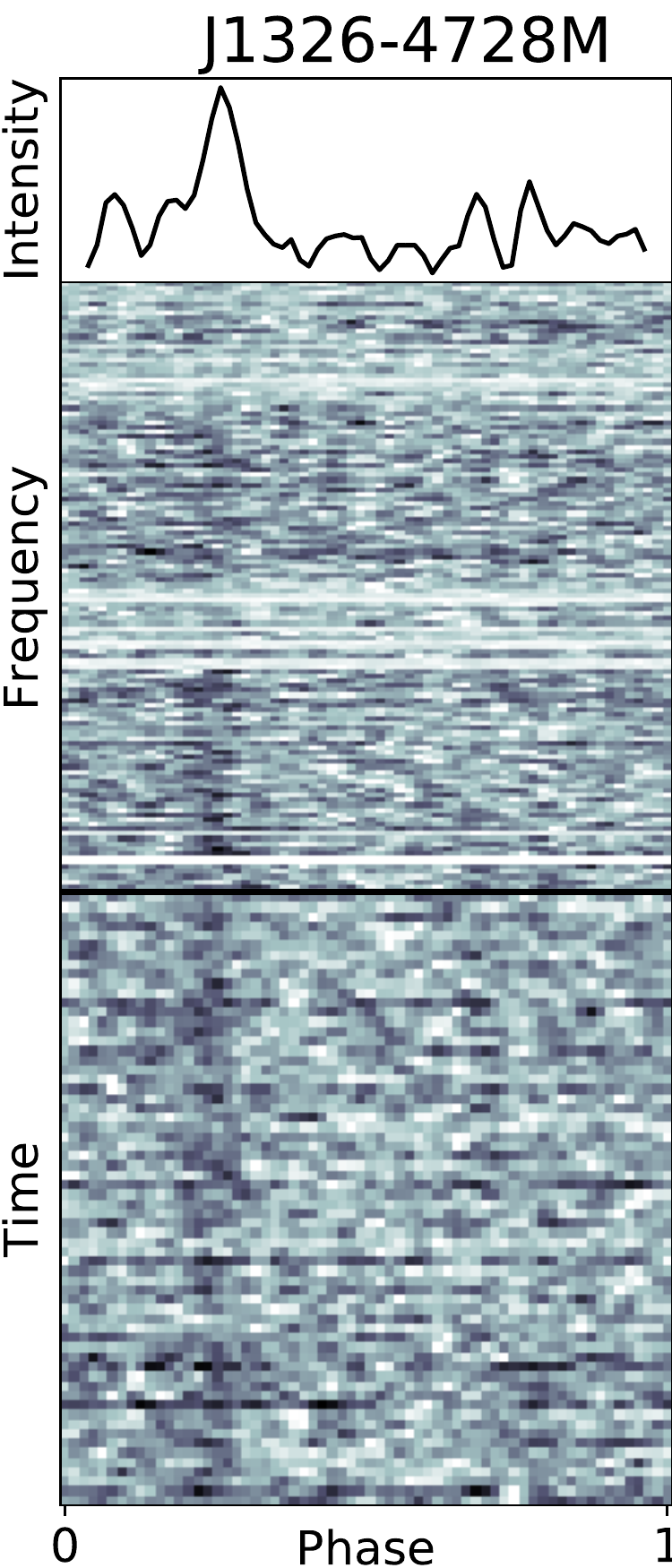} &
    \includegraphics[width=0.14\textwidth]{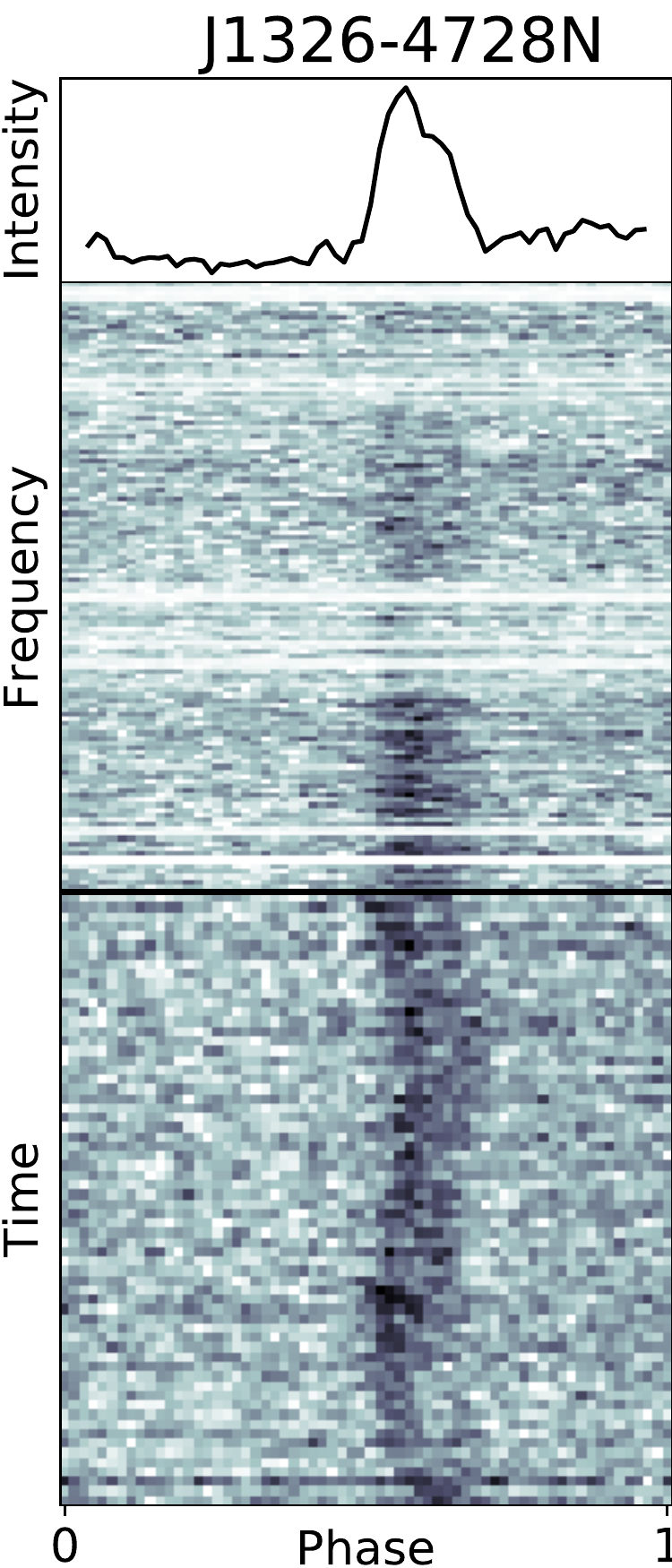} &
    \includegraphics[width=0.14\textwidth]{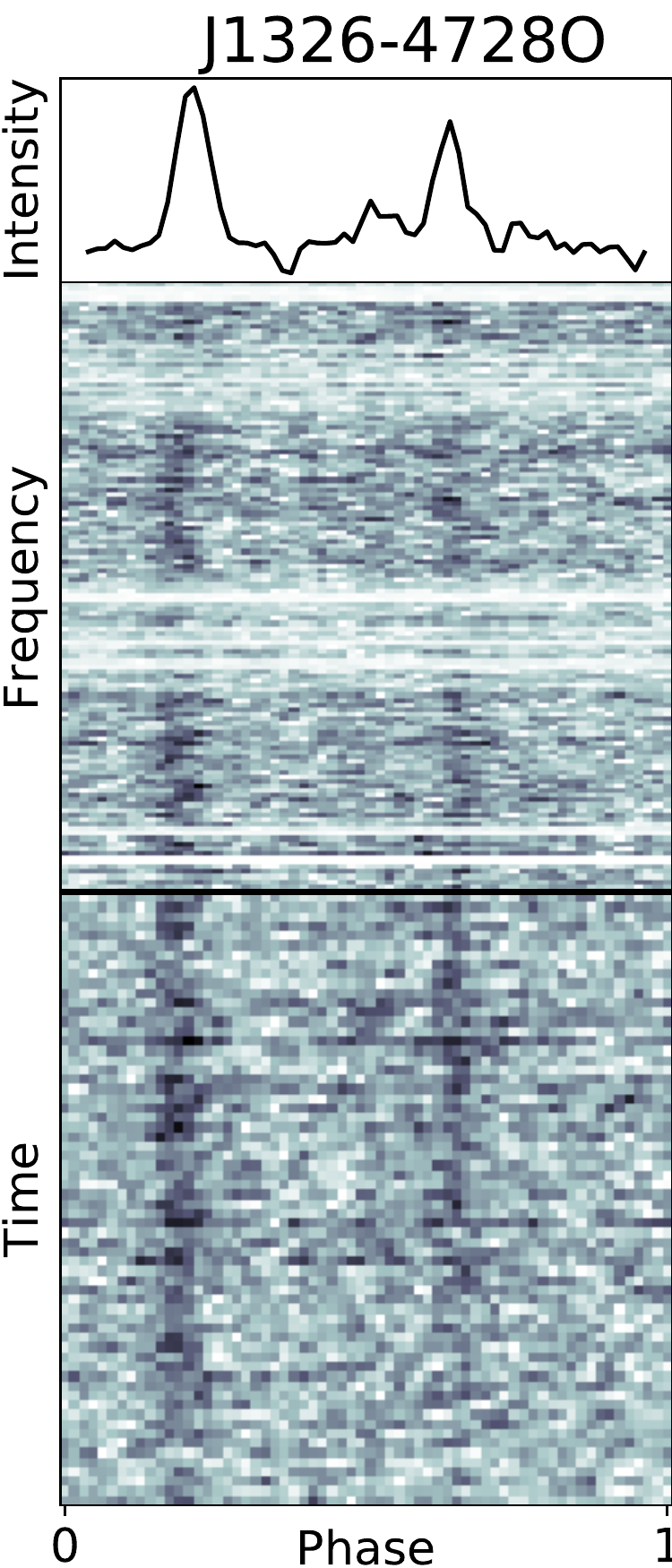} &
    \includegraphics[width=0.14\textwidth]{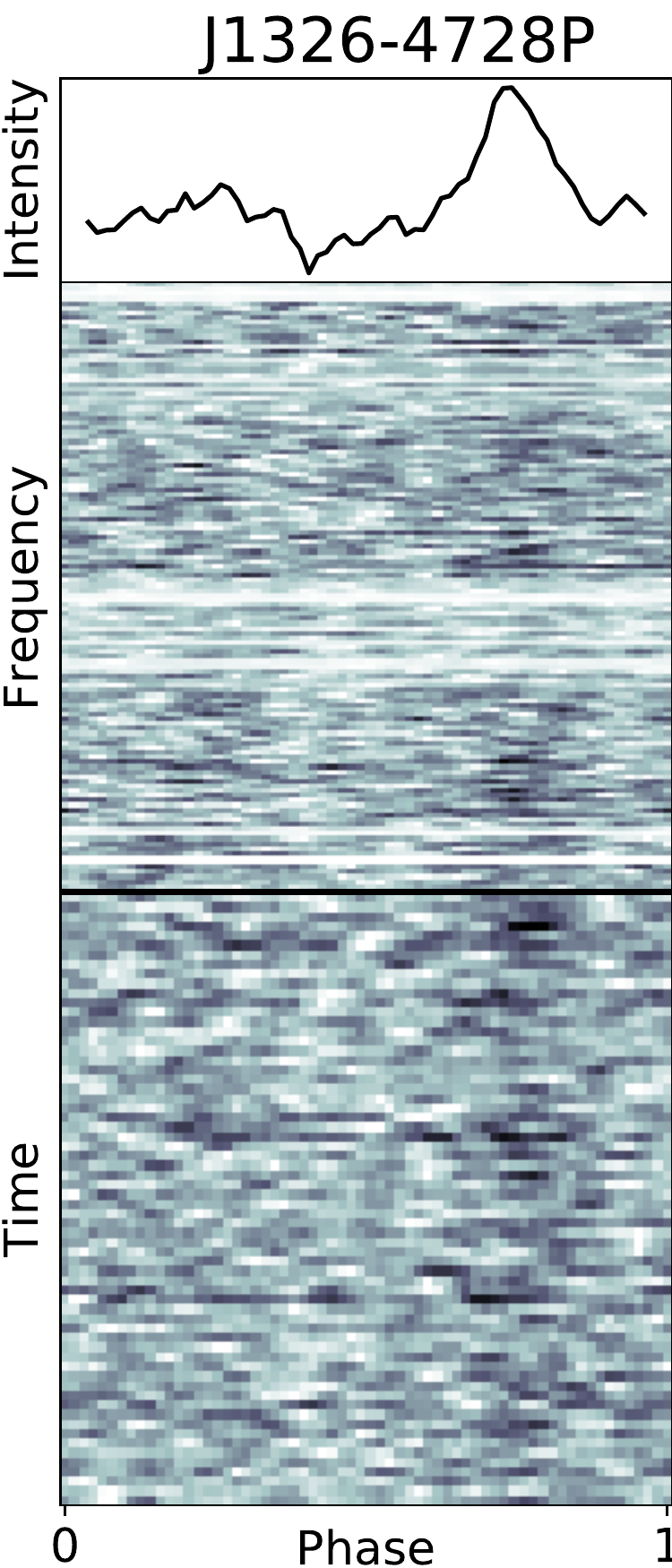} &
    \includegraphics[width=0.14\textwidth]{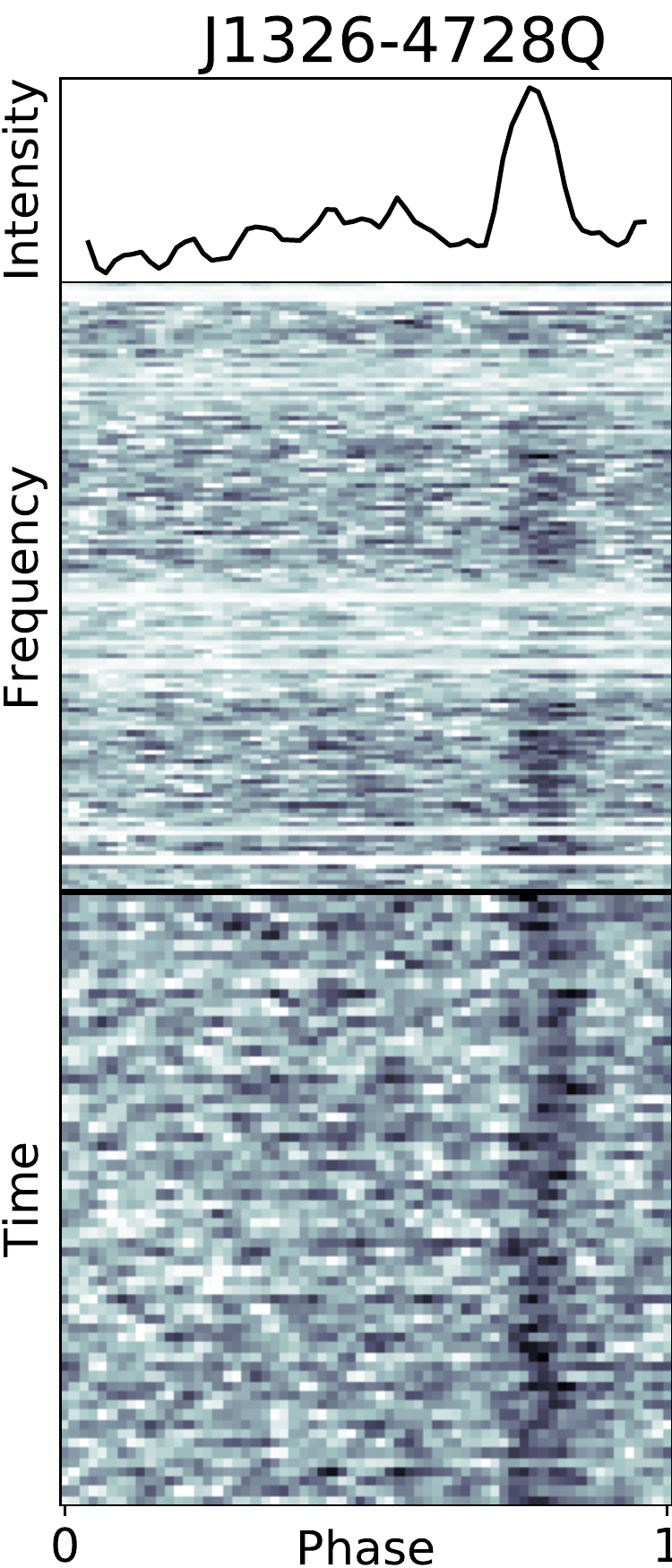} &
    \includegraphics[width=0.14\textwidth]{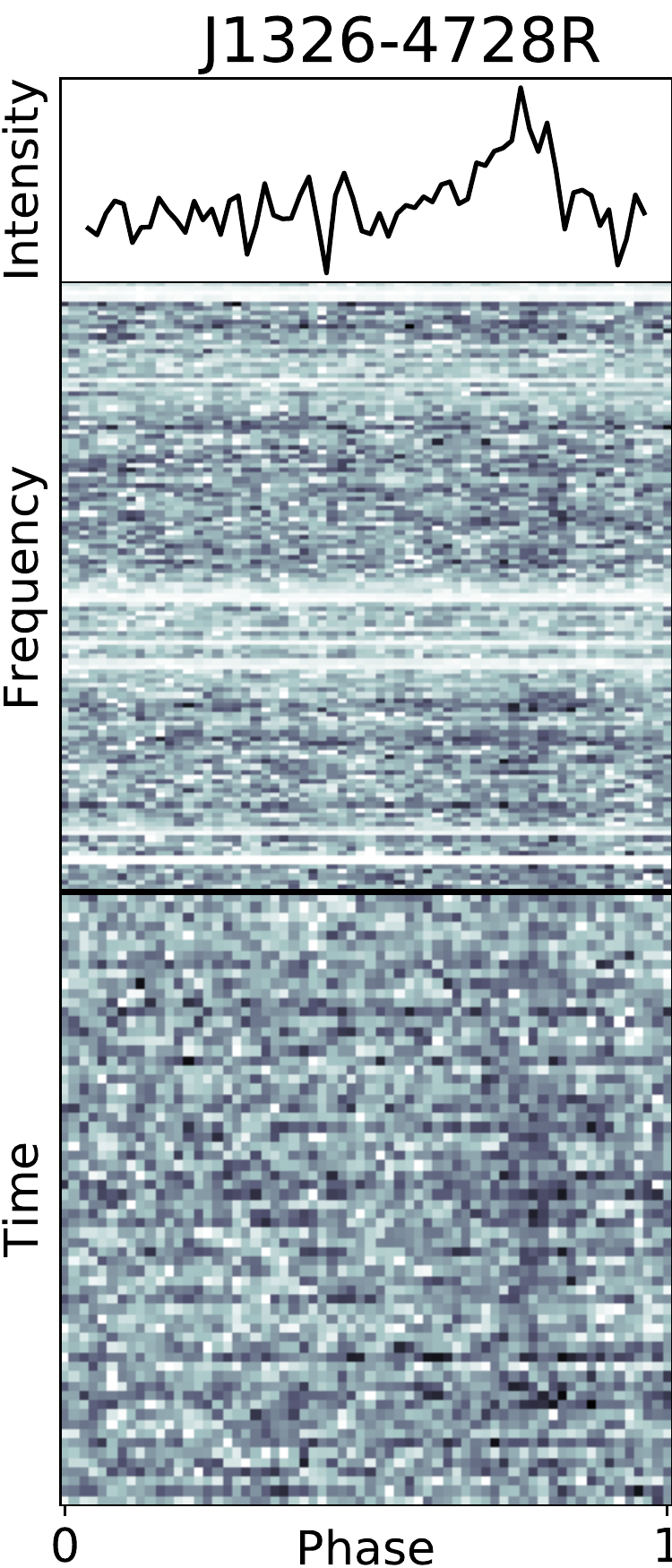}
    \end{tabular}
    \caption{Profiles of the new pulsars folded with 4 hours of data taken on 26 March 2021. The Y-axis is intensity, frequency and time from top to bottom panels, and X-axis is the phase window from 0 to 1.}
    \label{fig:profiles_of_new_pulsars}
\end{figure*}

\begin{figure}
    \centering
    \includegraphics[width=0.4\textwidth]{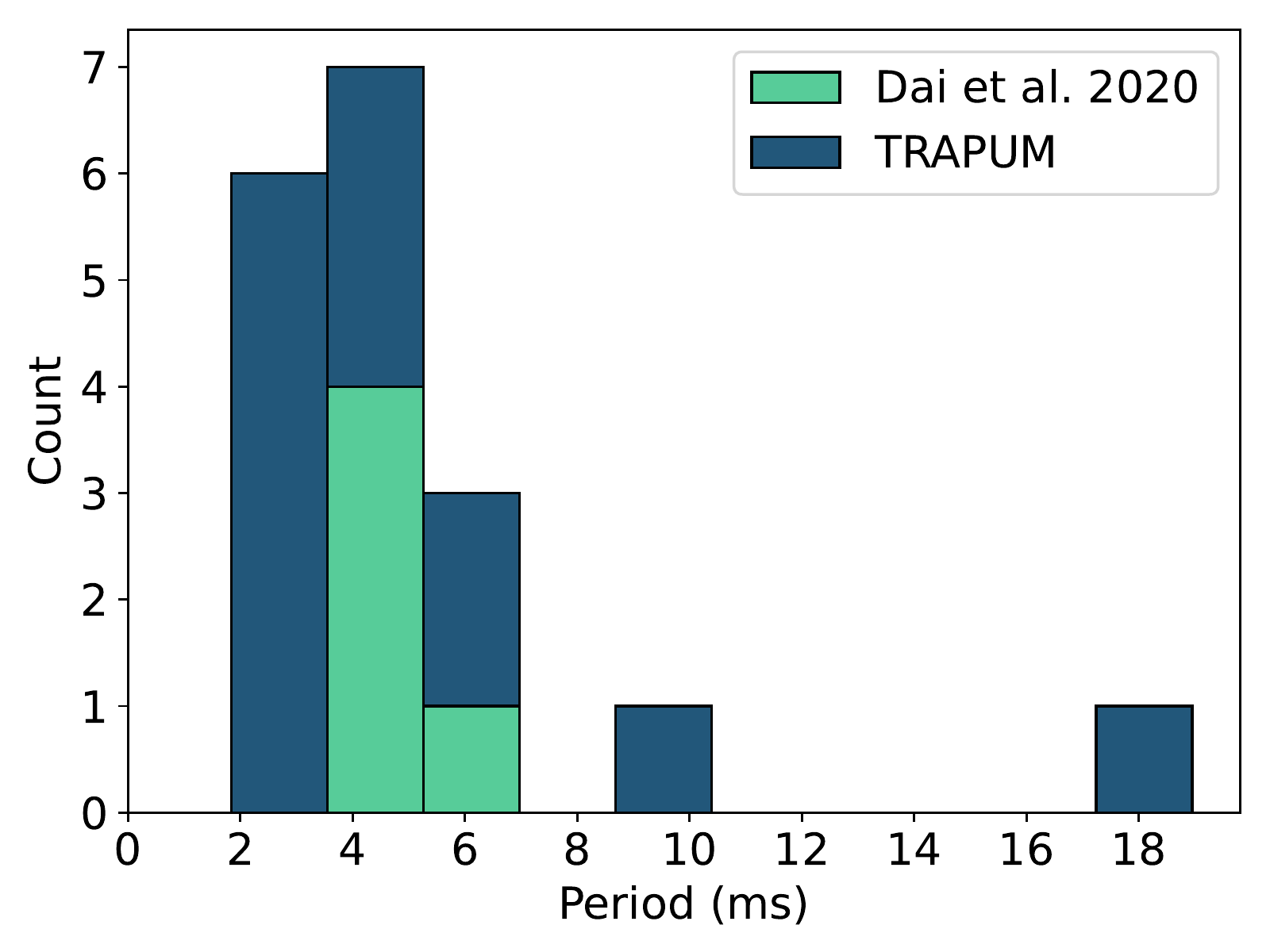}
    \caption{Distribution of spin period of pulsars in $\omega$-Cen.}
    \label{fig:spin_period_distribution}
\end{figure}

Our search resulted in the discovery of 13 new pulsars in the beams that were analyzed in this work (shown with blue edges in Figure \ref{fig:tiling}). Seven of them are in binary systems and we are able to place orbital constraints based on two observations. The results show that their orbital periods fall into two groups: below 4 hours and around 1 day. Notably, all except one have light companions. Other than the new discoveries, we also re-detected all the previously known pulsars \citep{Dai2020} in the cluster. The properties of all the new pulsars and which segment they were discovered, are listed in Table \ref{tab:discovery_table}, while their integrated profiles are shown in Figure \ref{fig:profiles_of_new_pulsars}. The distribution of their spin periods are presented in Figure \ref{fig:spin_period_distribution}. In the remainder of the section, we discuss their characteristics in more detail.

\begin{table*}
    \caption{List of discoveries from this work and their properties. The DMs were give by \texttt{prefold}, the orbital parameters were derived using \texttt{tempo}, the positions were localized using \texttt{SeeKAT} and the numbers in parentheses of the positions represent 2-$\sigma$ uncertainty of the last digit. The search lengths are the length of segments which the pulsars were discovered (see section \ref{subsec:data_reduction_and_saerch_pipeline}). }
    \begin{tabular}{c|c|c|c|c|c|c|c|c|c|c|c}
         \hline \hline
         &  Pulsar & Type     &    $P$    &  DM            & $P_b$ & $x_p$  & $M^{\textrm{min}}_c$ & $\alpha$ & $\delta$ & Search length \\
         &         &          &   (ms)  & (pc cm$^{-3}$) &  (d)  & (lt-s) & M$_\odot$  & J2000 & J2000 &   (h)  \\  \hline
         &  F      & Isolated &   2.27 &  98.29  &   -   &   -      &   - & $13\h26\m53\s(1)$ &  $-47\degr28\arcmin28\arcsec(6)^{\ddag}$ & 4 \\
         &  G      & Binary   &   3.30 &  99.69  &  0.1087597(1)  &  0.032203(7)  &  0.018  & $13\h26\m37\s.1(2)$ &  $-47\degr29\arcmin41\arcsec(1)$ & 0.5 \\
         &  H      & Binary   &   2.52 &  98.09  &  0.1356948(5)  &  0.021844(9)  &  0.011 & $13\h26\m44\s(1)$ &  $-47\degr28\arcmin55\arcsec(4)$ & 0.5 \\
         &  I      & Binary   &   18.95 &  102.2  &  1.113(1)$^{\flat}$   &  0.165(1)   &  0.020 & $13\h26\m29\s.0(1)$ &  $-47\degr30\arcmin24\arcsec(1)$ & 4   \\
         &  J      & Isolated &   1.84 &  97.28  &   -   &   -   & -    & $13\h26\m51\s.7$(1) &  $-47\degr27\arcmin09\arcsec(1)$ & 4  \\
         &  K      & Binary   &   4.72 &  94.73  &  0.09387146(2)  &  0.067945(5)  &  0.043 & $13\h26\m38\s.1(1)$ &  $-47\degr27\arcmin39\arcsec(2)$ & 0.5  \\
         &  L      & Binary   &   3.54 &  101.5  &  0.1589282(4)   &   0.061809(9)  &  0.027 & $13\h27\m02\s.8(1)$ &  $-47\degr26\arcmin49\arcsec(2)$ & 0.5 \\
         &  M      & Isolated &   4.60 &  101.4  &   -   &   -   & -     & $13\h26\m59\s(1)$ &  $-47\degr30\arcmin09\arcsec(6)^{\ddag}$ & 4 \\
         &  N      & Binary   &   6.88 &  101.2  &   1.0816(3)$^{\flat}$   &  0.1250(3)  &   0.015 & $13\h26\m49\s.8(1)$ &  $-47\degr31\arcmin25\arcsec(1)$ & 4 \\
         &  O      & Isolated &   6.16 &  94.27  &   -   &   -   & -     & $13\h26\m48\s(1)$ &  $-47\degr27\arcmin19\arcsec(6)^{\ddag}$ & 4 \\
         &  P      & Isolated &   2.79 &  102.1  &   -   &   -   & -     & $13\h26\m45\s(1)$ &  $-47\degr29\arcmin42\arcsec(6)^{\ddag}$ & 4 \\
         &  Q      & Binary   &   4.13 &  95.92  &  1.18(8)$^{\flat}$  &  1.1(1)  &   0.138 & $13\h26\m35\s(1)$ &  $-47\degr27\arcmin54\arcsec(6)^{\ddag}$ & 2 \\
         &  R      & Isolated &   10.29 &  102.1  &   -   &   -   &  - & $13\h27\m10\s(1)$ &  $-47\degr29\arcmin02\arcsec(6)^{\ddag}$ & 4 \\
         \hline
    \end{tabular} \\
    \begin{flushleft}
    \textbf{Note}: \\
    $^{\flat}$The orbits of these pulsars are longer than the observation, the values shown here are derived from two observations, hence they are subject to change when more data are available. \\
    $^{\ddag}$ The position of these pulsars was denoted by the centers of the beams where they are detected with highest S/N. Because of the lack of or faint detection in their neighbouring beams, further localization is not practical with \textsc{SeeKAT}.
    \end{flushleft}
    \label{tab:discovery_table} 
\end{table*}

\subsection{Isolated pulsars}

% J1326−4728F
% J1326−4728J
% J1326−4728M
% J1326−4728O
% J1326−4728P
% J1326−4728R

From our two observations, PSRs J1326$-$4728F, J, M, O, P and R appear to be isolated, because they do not show changes to their barycentric period to within the limits of the uncertainty. Most of them are within the core radius of the cluster (located at 2.37 arcmin from the nominal center of $\omega$-Cen), while pulsar M is near the edge of the core and pulsar R is between the core and half light radius (at 5.00 arcmin). Among them, J1326$-$4728J has a period of 1.84 ms, making it the fastest spinning pulsar so far discovered in this cluster. It also has a large duty-cycle of 43.7\%. For J1326$-$4728O and P, there were harmonic detections with similar S/N, it is ambiguous as to the number of peaks in the profile and thus the spin period is also ambiguous, further observations with polarimetric capability should resolve these ambiguities. The isolated pulsars discovered in these two observations are relatively weak compared to the re-detections of the known pulsars discovered in \cite{Dai2020}, suggesting that the sensitivity could be one of the reasons why the new pulsars were not detected in the previous observations with Parkes. 

\subsection{Binary pulsars}

\begin{figure}
    \centering
    \includegraphics[width=0.4\textwidth]{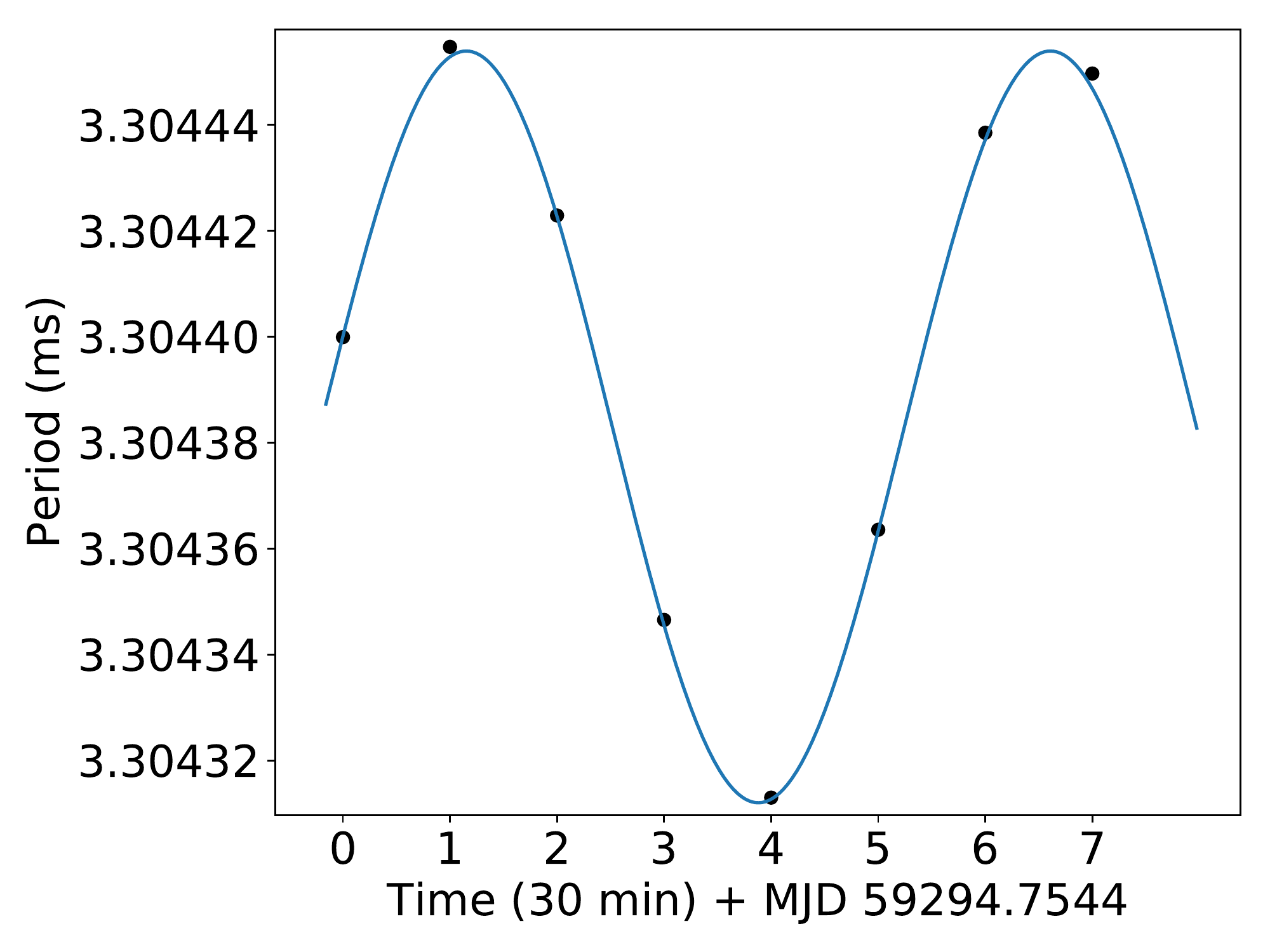}
    \caption{Changes of periods of J1326$-$4728G during the first four-hour observation. The data points are 30- minute segments and the curve is an orbital fit of the period changes given by \texttt{fit\_circular\_orbit.py}.}
    \label{fig:orbit_G}
\end{figure}

More than half of the discoveries in this work reside in binaries. To obtain their orbital parameters, the data were first split into segments, from which time-dependent barycentric period and period derivative estimates could be made. An example of such practice for J1326$-$4728G is shown in Figure \ref{fig:orbit_G}. With these segmental sets of values, the orbital parameters were fit assuming a circular orbit, using the \texttt{fit\_circular\_orbit.py} Python script from \textsc{presto}. These solutions were then used as the initial guess for the orbital parameter when constructing an initial ephemeris for each binary pulsar. Subsequently, these ephemerides were iteratively improved by pulsar timing using the \textsc{tempo}\footnote{\url{https://sourceforge.net/projects/tempo}} software package. According to the current ephemerides, pulsar J1326$-$4728I, N and Q have orbits longer than the observation, wherefore, the solution of these orbits are not unique and subject to change where more data are available. 

% J1326−4728G
% J1326−4728H

J1326$-$4728G and H have periods of 3.30 and 2.52 ms, respectively. The former was found near the edge of the core and was detected brightly in both epochs and in neighbouring beams, the latter was found near the centre of the cluster. Preliminary orbital solutions suggest that they have orbits of approx. 2.61 hours and 2.36 hours, respectively. According to the mass function and assuming the mass of the pulsar is 1.4 M$_\odot$ (also for the following paragraphs), the mass range of the companion of J1326$-$4728G is [0.018, 0.042] M$_\odot$ when the inclination angle between the orbital plane and the light of sight is $0\degr$ and $64\degr$. The upper bound was chosen such that the range covers 90\% of the cases \citep{LorimerKramer2004}, assuming the orbital planes are isotropically distributed. Similarly, the mass range of the companion of J1326$-$4728H is [0.011, 0.024] M$_\odot$. The orbital periods and minimum companion masses fall within the typical range of expected values for ``black widow'' systems \citep{Roberts2013}. Though, no eclipses, which are often seen in such systems, are observed in these two observations. But considering the tight orbits, low mass companions and rapid spin periods, it is very likely they are black widows.

% J1326−4728K
% J1326−4728L
J1326$-$4728K and L have periods of 4.71 and 3.53 ms, respectively. The former was discovered close to the edge of the core and was brightly detected in both epochs and in neighbouring beams, the latter was discovered between the radii of the core and the half light. An interpulse can be clearly observed in the profile of J1326$-$4728K. Fits to the detected spin period and spin period derivatives suggest orbits of about 2.25 and 3.81 hours, with the ranges of companion masses of [0.043, 0.101] M$_\odot$ and [0.027, 0.063] M$_\odot$, respectively. Again, these are values typical of black widow systems. For pulsars K and L, however, eclipses are clearly seen in both observations.

% J1326−4728I
J1326$-$4728I has a spin period of 18.95 ms with a relatively large duty cycle of 29.7\%. It is found between the radius of the core and of the half light of the cluster. The profile can be phase-aligned well, assuming a constant line-of-sight acceleration in each observation. However, the sign of the acceleration changed between two observations. This suggests that the pulsar is moving in a binary system with an orbit that is significant longer than the length of our observations. The orbital solution suggests that it has an orbit of 26.71 hours and its companion has a mass range of [0.020, 0.046] M$_\odot$. Additionally, the period of the pulsar indicates it might be a mildly recycled pulsar. But without an accurate period derivative, other binary evolution scenario cannot be ruled out. Further observations could shed light on the nature of this pulsar.   

% J1326−4728N
% J1326−4728Q
J1326$-$4728N and Q have periods of 6.88 and 4.13 ms. They were both discovered at the edge of the core. Their orbital parameters suggest orbits of about 25.96 and 28.41 hours, with mass ranges of their companions of [0.015, 0.035] M$_\odot$ and [0.138, 0.342] M$_\odot$, respectively. It is difficult to constrain the orbital parameters for J1326$-$4728N, which could be partially due to the low and uneven orbital phase coverage during two observations. Pulsar Q was brightly detected in one observation, but could not be re-detected in the expected beam in the other observation. There are, however, very faint detections in two of the neighbouring beams of that nearest beam. The companion mass of Q indicates that it could be a helium white dwarf. Follow-up observations are crucial for improving the orbital solutions of these two pulsars.

\subsection{Localization and correlation with X-ray emission}
\begin{figure}
    \centering
    \includegraphics[width=0.37\textwidth]{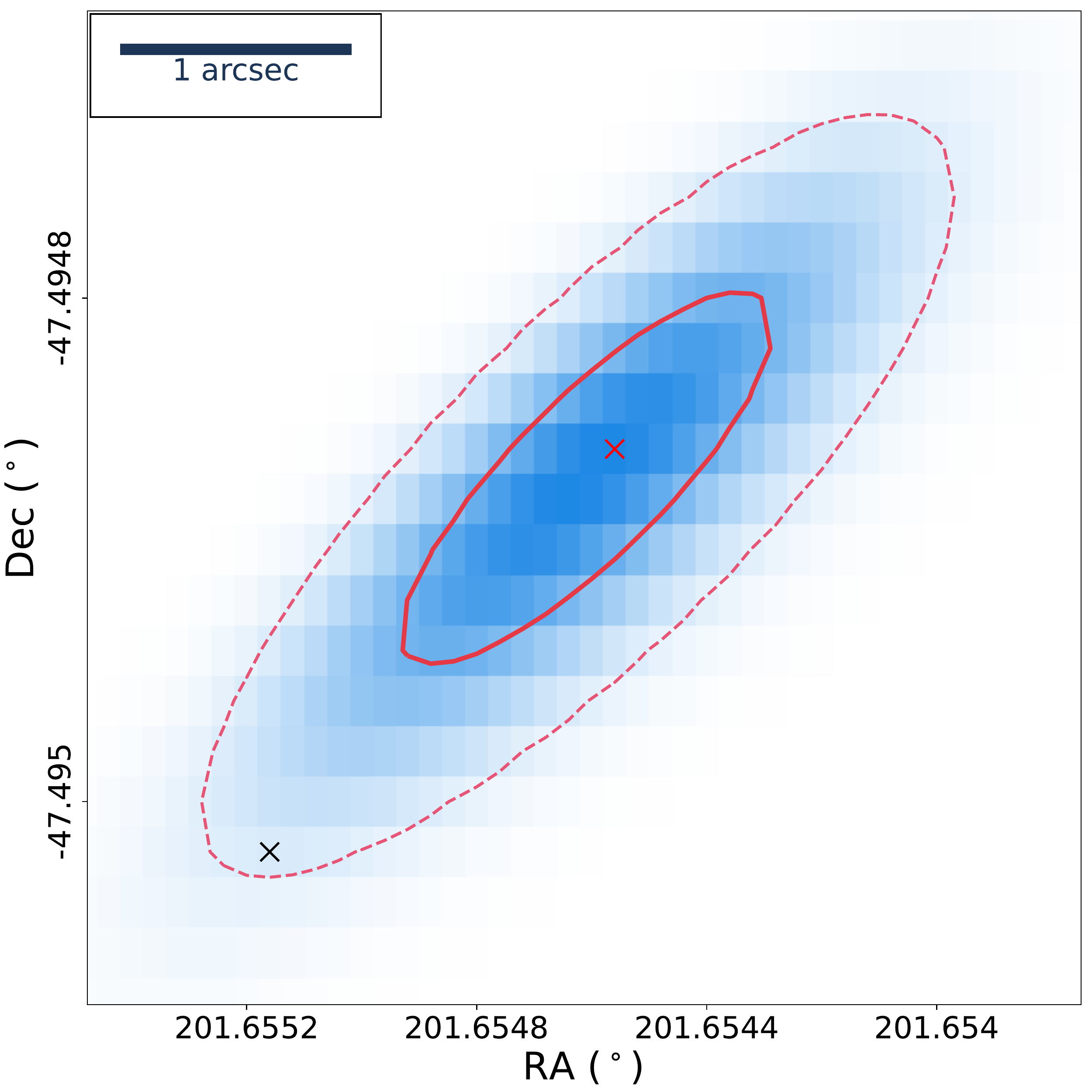}
    \includegraphics[width=0.37\textwidth]{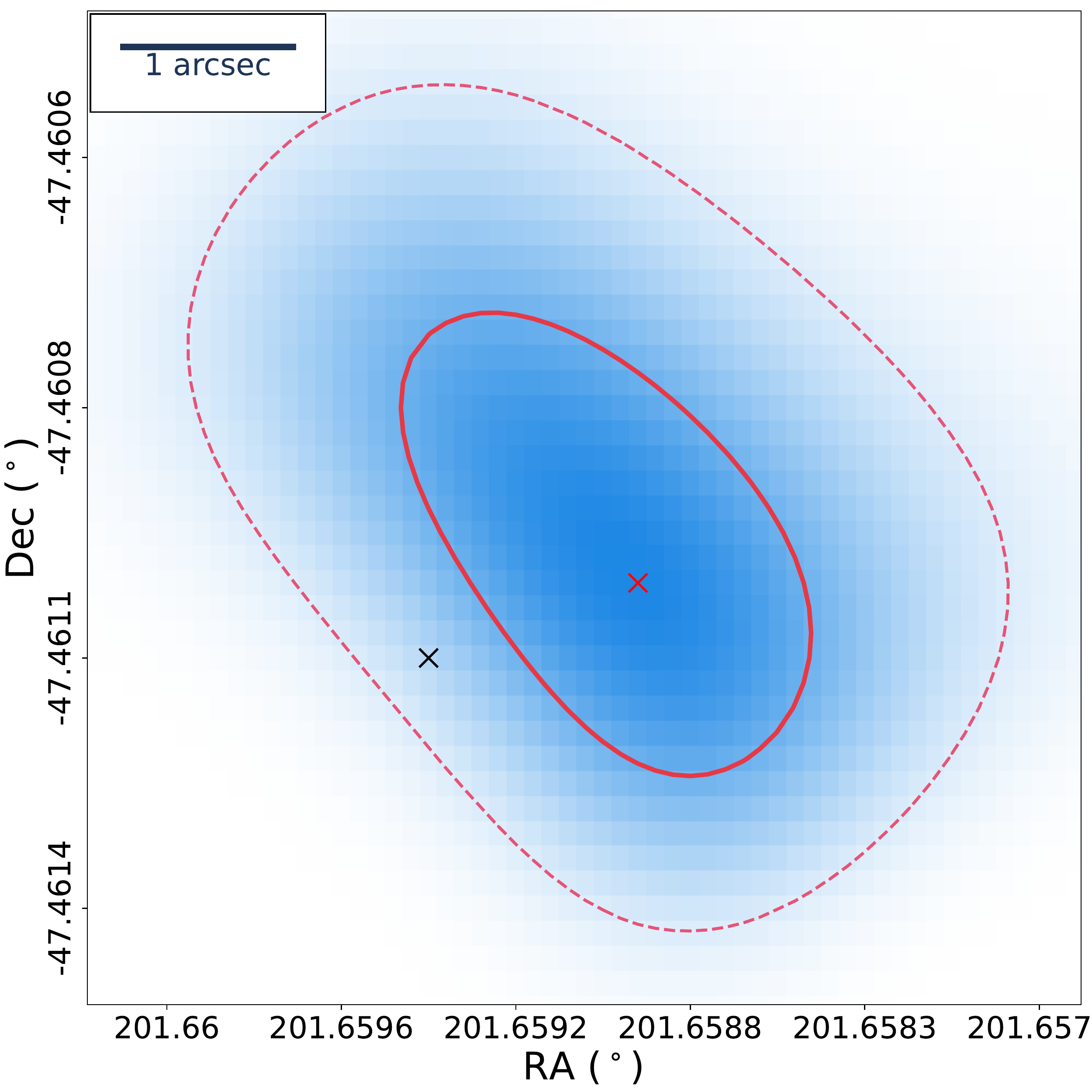}
    \caption{Localization and nearby X-ray sources. Top: localization result of J1326$-$4728G, red cross is the best position given by \textsc{SeeKAT}, black cross is the position of the nearest X-ray source \texttt{24f} from \citet{Henleywillis2018}. Blue shades are the likelihood map of the position. The solid line and dashed line are 1-$\sigma$ and 2-$\sigma$ confident levels of the result. Bottom: localization result of J1326$-$4728K and the position of the nearest X-ray source \texttt{21d}.}
    \label{fig:localization_plots}
\end{figure}

About half of the new pulsars were detected in multiple neighbouring beams, allowing their positions to be constrained to a precision better than the size of the coherent beam. The localization was carried out using the \textsc{SeeKAT}\footnote{\url{https://github.com/BezuidenhoutMC/SeeKAT}} multibeam localization software  (Bezuidenhout et al., submitted). It performs maximum-likelihood estimation to obtain the best position. The likelihood is calculated by testing the theoretical gain of the position given by a constant point spread function (PSF), against the S/N's of the neighbouring detections. The PSFs were generated using \textsc{mosaic} with the observational parameters. \par
Our two observations lasted four hours each, and the PSF changed with time. Significant changes of the PSF would lead to deterioration of localization quality, therefore we tried to mitigate this effects by using the S/N's and PSFs in different segments of the observations. For faint pulsars, it is not practical to obtain detections in short segments, in those cases, the PSFs corresponding to the middle of the observation were used. The result of the localization were shown in Table \ref{tab:discovery_table}. Apart from the new pulsars, we also perform localization with the known pulsars, because three of them have no timing position at the moment. The results for the known pulsars are listed in Table \ref{tab:know_pulsar_positions}. Here, we present the localization of J1326$-$4728G and K plotted with X-ray sources (24f and 21d) within the 2-$\sigma$ confident levels shown in Figure \ref{fig:localization_plots}. It is worth noting that multibeam localization relies strongly on the accuracy of the PSF and the quality of the data, such as errors on the S/N, severity of the RFI, coherence of the signal etc., all of which were not considered in this estimation. \par
Previous observations made with the Parkes telescope \citep{Dai2020} associated one X-ray source from \citet{Henleywillis2018} to PSR J1326$-$4728B. However, there are many X-ray sources in this cluster that are still unassociated. Hence, we compared them with the new radio pulsar discoveries and noticed some of the new pulsars have nearby X-ray sources within a few arcseconds, such as the binaries pulsars J1326-4728G, H, K and L. However, it is difficult to deduce a firm association until we obtain more accurate radio timing position from follow-up observations. Since we now know that this cluster hosts a considerable amount of pulsars, it is likely that some of the X-ray emission from this cluster comes from pulsars. \par
 
\begin{table}
    \caption{Positions of the known pulsars localized with \textsc{SeeKAT}. The numbers in parentheses represent 2-$\sigma$ uncertainty of the last digit.}
    \centering
    \begin{tabular}{c|c|c}
         \hline \hline
         Pulsar & $\alpha$     & $\delta$  \\
                & J2000        & J2000     \\ \hline
         A & $13\h26\m39\s.7(1)$ & $-47\degr30\arcmin11\arcsec(2)^{\ddag}$ \\
         B & $13\h26\m49\s.7(1)$ & $-47\degr29\arcmin26\arcsec(2)^{\ddag}$ \\
         C & $13\h26\m55\s.5(1)$ & $-47\degr30\arcmin13\arcsec(2)$ \\
         D & $13\h26\m32\s.5(1)$ & $-47\degr28\arcmin39\arcsec(4)$ \\
         E & $13\h26\m42\s.6(1)$ & $-47\degr27\arcmin22\arcsec(1)$ \\
         \hline
    \end{tabular}
    \begin{flushleft}
    \textbf{Note}: \\
    $^{\ddag}$ J1326$-$4728A and B have timing positions from \citet{Dai2020}, which are $13\h26\m39\s.670, -47\degr30\arcmin11\arcsec.64$ and $13\h26\m49\s.563, -47\degr29\arcmin24\arcsec.62$.
    \end{flushleft}
    \label{tab:know_pulsar_positions}
\end{table}

\section{Discussion}
\label{sec:discussion}

\subsection{The pulsar population in $\omega$-Cen}
\label{subsec:pulsar population}

The number of pulsars known in $\omega$-Cen, which was 5 before this work, is now 18. 
Taking the characteristics of the previously known pulsars, plus those of Table~\ref{tab:discovery_table} at face value, we see that the pulsar population of $\omega$-Cen appears to be dominated by isolated pulsars. Indeed, the previously known isolated pulsars A, C, D, E plus the newly discovered isolated pulsars, F, J, M, O, P and R represent 10 out of a total of 18 pulsars. Furthermore, we can also see that, with the exception of Q, all binary pulsars have very low-mass companions, typical of what one finds in ``black widow'' systems. It has been noted that black-widow pulsars are more easily formed in GCs \citep{King2003}, but their very high faction in $\omega$-Cen is still surprising. Of these, three systems (I, N and Q) seem to have orbital periods of just over 1 day, the other four have orbital periods smaller than the length of one observation, 4 hours.

However, before we advance with a detailed analysis of the characteristics of this population, it is important to keep in mind several caveats.

First, about the percentage of isolated pulsars, the low stellar density of $\omega$-Cen means that wide binary systems with orbital periods of tens or even hundreds of days might be stable in this cluster (as, for example, PSR B1310+18A, a 255-day, low-eccentricity binary in the globular cluster M53, see \citealt{Kulkarni1991}). This suggests that some of the pulsars that are apparently isolated might, with additional observations, be found to be part of wide binary systems, where the apparent changes of spin period caused by the orbital motion are less pronounced.

Second, as mentioned above, we have analyzed in our search several stretches of data going up to 4 hours. Finding binaries in such long integrations is more difficult than finding isolated pulsars, because of the loss of sensitivity caused by the orbital motion; this is especially true for cases where the total integration time is of the same order as the orbital period. 
This is the reason why we find, from the last column in Table~\ref{tab:discovery_table}, that while all isolated pulsars and longer period binaries were discovered in 4-h segments (with the exception of Q, which was found in a 2-h segment), no short-period binaries were found in such long segments: they were instead found, invariably, in 30-minute segments.
This means that, in this survey, we are $\sqrt{4/0.5} = 2.83$ times more sensitive to very faint isolated MSPs (or MSPs with very wide obits) than to short binaries. This means that the isolated pulsars and binary pulsars with low accelerations are over-represented in our sample relative to the short-period binaries.

Third, for normal MSP - WD binaries with orbital periods of a few days, no orbits can be firmly determined with only two observations; to determine their orbits additional observations will be necessary. We note in this regard that the 1-day orbits for I, N and Q are preliminary, thus these  binary MSPs could in principle have more massive companions and longer orbital periods.

Summarizing, and taking these caveats into account, it does appear that a) the number isolated MSPs (10 out of 18) is large, but represents an over-estimation of the pulsar population of the cluster, because they are easier to find and also because some of them could potentially be wide binaries; b) the confirmed short-orbit systems (those with short orbital periods: B, G, H, K and L) are more numerous than the other confirmed longer-period binaries (I, N and Q); their numbers are likely to be under-estimated, since we are less sensitive to these short-orbit binaries; c) The longer-period binaries might not be fully characterized yet, they could have larger orbital periods and more massive companions; two of them (I and N) could still have very light companions.

From this, we can conclude firmly that the fraction of black widow systems in $\omega$-Cen is unusual. The only comparable GC is M28, where they represent half of the total binary population \citep{Douglas2022}. However, M28 has vastly different properties, in particular a much denser core. In that GC, almost all long-period binaries have been either disrupted, or show significant eccentricities or even possible signs of having undergone secondary exchange encounters. Only the BWs survive because their very short orbital periods make them much more difficult to perturb: they are smaller targets and require a more energetic encounter for the orbit to change significantly.

\subsection{Why this pulsar population is surprising}

The total stellar encounter rate ($\Gamma$) and the stellar encounter rate per binary ($\gamma$) are functions of the core radius ($r_c$) and central density ($\rho_c$) of the GCs, with $\Gamma \propto \rho_c^{1.5} r_c^2$ and $\gamma \propto \rho_c^{0.5} r_c^{-1}$ \citep{Verbunt1987,Verbunt2014}. The first parameter, $\Gamma$, gives an approximate prediction of the size of the pulsar population, the second gives an approximate prediction of how disturbed the binaries are: in GCs with low $\gamma$, the population resembles the MSP population in the Galactic disk. As $\gamma$ increases, more frequent encounters have a higher chance of perturbing the orbits of the binaries and on disrupting systems, producing more eccentric binaries and more isolated pulsars overall. Finally, for core-collapsed clusters with 5 or more pulsars known, such as Terzan 1, NGC 6517, NGC 6522, NGC 6624, NGC 6752 and M15, the population is completely dominated by isolated pulsars, while a large percentage of binaries results from secondary exchange encounters (for recent discussions, see \citealt{Ridolfi2021,Ridolfi2022}).

$\omega$-Cen has a low stellar density compared to most other GCs ($10^{3.15} \rm L_\odot pc^{-3}$), which results in rather low values for $\Gamma$ and $\gamma$ of respectively 4.3 and 0.11 (these are normalized to the values of the GC M4, as in \citealt{Verbunt2014}, as in that work we have used the values of $r_c$ and $\rho_c$ from \citealt{Harris2010}). As a comparison, 47~Tuc has $\Gamma = 29.2$ and $\gamma = 6.6$.

These $\Gamma$ values mean that LMXBs and MSP binaries should form in $\omega$-Cen at a rate $\sim7$ times smaller than that of 47~Tuc. However, the 18 pulsars in $\omega$-Cen are two thirds of the 27 pulsars detected in 47~Tuc with the same telescope (MeerKAT) and receivers (L-band, see \citealt{Ridolfi2021}). This is partly a result of the fact that the latter searches were done only with a single beam using 44 antennas within the 1-km core, not all 64 antennas and many hundreds of beams as the search described here; however this difference is in part compensated by the larger distance of $\omega$-Cen (5.2 kpc) compared to 47~Tuc (4.5 kpc). In any case, $\omega$-Cen is surprisingly effective in producing MSPs.

The low $\gamma$ means that, once formed, there is not much chance of a significant disruption of these systems: indeed, the interval between successive interactions with other stars should be $\sim 60$ times larger in $\omega$-Cen than in 47~Tuc. In the latter cluster, we see a MSP population that resembles the pulsar population of the Galactic disk (except for the absence of long-period binaries, see \citealt{Ridolfi2016,Freire2017}). Therefore, the pulsar population of $\omega$-Cen should also be similar to the MSP population in the Galactic disk, which is dominated by binaries, in an approximate rate of 4 to 1. As discussed above, the fraction of isolated pulsars in $\omega$-Cen appears to be large, but this might be mostly due to selection effects, so it is still possible that the pulsar population of $\omega$-Cen is dominated by binaries. 

More difficult to explain is the small population of MSP-WD systems and long-period binaries. As discussed above, the predominance of very tight systems is likely real, and it is not something that can be expected from the dynamical characteristics of this GC.

It is therefore clear that the properties of the pulsar populations of $\omega$-Cen cannot be fully explained by two simple parameters like $\Gamma$ and $\gamma$: despite selection effects, we can already conclude that this cluster has a larger pulsar population than expected from its $\Gamma$, too many black widow systems and possibly too many isolated pulsars. That $\Gamma$ and $\gamma$ do not provide a full description of pulsar populations in GCs was already highlighted by \cite{Verbunt2014} when they compared the pulsar populations of NGC 6440/1 with that of Terzan 5, which has a similar $\Gamma$ and $\gamma$: the spin period distribution of the pulsars in both populations is completely inconsistent. Therefore, the characteristics of the pulsar populations in GCs must depend on additional factors, like the past evolutionary history of the GCs and potentially their metallicity.

In the specific case of $\omega$-Cen, the past history of the system might be of paramount importance. For instance, $\omega$-Cen could have been the nucleus of a dwarf galaxy \citep{Hilker2000,Bekki2003}, alternatively, it might have formed from the merger of several different GCs \citep{Calamida2020}. Such explanations are motivated by the fact that this GC has multiple stellar populations, with different metallicities and ages. Given the typical ages of MSPs (many Gyr), the dramatic events in the history of these GCs should be of paramount importance for an explanation of the characteristics of its pulsar population today.

We note in this regard that there are other GCs in the Galaxy that are known to have multiple stellar populations, and are likely associated with dwarf galaxy systems or are the results of GC mergers. Two of the most prominent are Terzan 5, where three distinct stellar populations have been found \citep{Ferraro2009,Origlia2013}, and NGC~1851 \citep{Carretta2011}. All have very abundant pulsar populations with about 50\% and 40\% of isolated pulsars respectively (e.g., \citealt{Ransom2005,Ridolfi2022}). However, their cores are so dense that the orbital characteristics of these pulsar populations have likely been significantly altered by exchange encounters. This is not the case for $\omega$-Cen, where the low $\Gamma$ and $\gamma$ mean that the orbital characteristics of the pulsar population have been preserved for a long time; they should therefore reflect the earlier evolutionary history of the cluster.

A detailed evaluation of these possibilities is beyond the scope of this work, but it will be a profitable exercise, especially after the pulsar population in $\omega$-Cen is better characterized.

\section{Summary and future prospects}

In this paper, we presented the discovery of 13 new pulsars in $\omega$-Cen, which more than tripled the population of known pulsars in this cluster. They are found within the core and also between the core and half light radius of the cluster. Among them, six are isolated pulsars and the other seven are binaries. More than half of the binaries have orbits less than 4 hours, which is the length of the observations; three other binaries have orbital periods of about 1 day, but confirming this will require additional observations. All but one of the binaries have very light companions and two of them have apparent eclipses.

Follow-up observations are crucial to improve the orbital parameters of the wide binaries (I, N and Q) and help estimate their companion masses, which is a first step to an accurate characterization of those systems. Additional observations will also be important for deriving phase-connected timing solutions. Thanks to the many beams systhesised in each observation, we were able to constrain the positions of several of the new pulsars and compare them with the position of X-ray unassociated sources; for some binaries, there is an X-ray source nearby, within a few arcseconds. With timing solutions, these positions will become orders of magnitude more precise, this will either confirm or rule out some of our preliminary associations. Multi-wavelength observations should be carried out to check if these sources are still emitting X-rays, in order to establish whether the emission comes from the pulsars themselves or from ongoing accretion. This kind of observations can also be used to investigate other X-ray sources that have no radio signals detected because it is possible that there is a LMXB system there and the pulsar is still accreting.
The timing solutions, with precise estimates of acceleration and proper motions, will also be important for characterizing the gravitational field of $\omega$-Cen \citep{Prager2017,Freire2017,Abbate2018}.

The large pulsar population, the large number of isolated pulsars and the fraction of black widow systems are surprising, considering the small encounter rate and the low encounter rate per binary of this GC. These parameters, although useful for an approximate characterization of the pulsar population of GCs, clearly do not tell the full story; it is very likely, for instance, that the past dynamical history of the GCs and the stellar evolution in binaries play important roles. This implies that the accurate characterization of the pulsar populations in GCs in general, and $\omega$-Cen in particular, provides valuable material for the study of stellar and cluster evolution. A particularly interesting possibility is that the pulsar population in $\omega$-Cen came from different smaller clusters that might have merged to form it \citep{Calamida2020}.

There are still more than half of the total beams outside the half light radius (5\arcmin) that have not been searched. Searching them will be important for finding out how centrally condensed the pulsar population of $\omega$-Cen is compared to other clusters. The Parkes survey by \cite{Dai2020}, which at L-band has a beam radius of 7.5\arcmin, has only found pulsars within, or very near the core, as discovered by our recent MeerKAT localisations and their subsequent pulsar timing. Our discoveries are also mostly within the core, with only five pulsars between 1 and 2 core radii from the centre of the cluster. The number of X-ray sources in $\omega$-Cen decreases significantly beyond 2 core radii, but still presents a detectable excess compared to the background beyond 3 core radii \citep{Henleywillis2018}. This suggests that additional pulsars might be detectable outside the half-light radius, but likely in significantly smaller numbers.

The `dynamical relaxation time'' in the core of $\omega$-Cen is 4~Gyr, while the median relaxation time for the cluster as a whole is 12~Gyr. For 47~Tuc, these numbers are 0.07 and 3.5~Gyr respectively, for Terzan 5, they are 0.037 and 0.34~Gyr respectively \citep{Harris2010}. What this  means is that, in 47~Tuc and Terzan 5, enough time has elapsed for mass segregation to occur, all pulsars in these two clusters (with the exception of 47~Tuc~X, \citealt{Ridolfi2016}) have moved to within 2\arcmin from their centres, and are likely in dynamical equilibrium with the remaining stars of the cluster (i.e., they are a ``relaxed" population, see e.g., \citealt{Heinke2005}). In $\omega$-Cen, this process takes much longer. This means that the current pulsar distribution, especially outside the core, likely reflects the ``original" dynamics of the pulsars within the cluster (either where they formed, or where they were placed by previous interactions of the cluster). A detailed dynamical study of this distribution could thus provide additional clues on the origin of this unusual pulsar population.

We also note that future TRAPUM observations with UHF-band (550-1100 MHz) and S-band (1750-3500 MHz) receivers will very likely further increase the population of known pulsars in $\omega$-Cen in all regions by probing different spectral windows.

\section*{Acknowledgements}

TRAPUM observations used the FBFUSE and APSUSE computing clusters for data acquisition, storage and analysis. These clusters were funded and installed by the Max-Planck-Institut für Radioastronomie and the Max-Planck-Gesellschaft.
WC, AR and FA acknowledge continuing valuable support from the Max-Planck Society.
LV acknowledges financial support from the Dean's Doctoral Scholar Award from the University of Manchester.
APo, AR and MBu gratefully acknowledge financial support by the research grant ``iPeska'' (P.I. Andrea Possenti) funded under the INAF national call Prin-SKA/CTA approved with the Presidential Decree 70/2016. APo, AR, MBu also acknowledge support from the Ministero degli Affari Esteri e della Cooperazione Internazionale - Direzione Generale per la Promozione del Sistema Paese - Progetto di Grande Rilevanza ZA18GR02.
The MeerKAT telescope is operated by the South African Radio Astronomy Observatory, which is a facility of the National Research Foundation, an agency of the Department of Science and Innovation. SARAO acknowledges the ongoing advice and calibration of GPS systems by the National Metrology Institute of South Africa (NMISA) and the time space reference systems department of the Paris Observatory. The National Radio Astronomy Observatory is a facility of the National Science Foundation operated under cooperative agreement by Associated Universities, Inc. SMR is a CIFAR Fellow and is supported by the NSF Physics Frontiers Center awards 1430284 and 2020265.
RPB acknowledges support ERC Starter Grant `Spiders' under the European Union’s Horizon 2020 research and innovation programme (grant agreement number 715051).

\section*{Data availability}
The data underlying this article will be shared on reasonable request to the TRAPUM collaboration.

\bibliographystyle{mnras}
\bibliography{references}

\end{document}